\pdfoutput=1

\documentclass[11pt,twoside,a4paper,cmspaper,final,collab]{cms-tdr}
\def\svnVersion{447294P}\def\cmsCernNoTag{CERN-EP-2017-252}\def\cmsCernDate{\today}\def\cmsMessage{Published in Physical Review D as \href{http://dx.doi.org/10.1103/PhysRevD.97.032009}{\doi{10.1103/PhysRevD.97.032009.}}}

\begin{document}\cmsNoteHeader{SUS-17-001}

\hyphenation{had-ron-i-za-tion}
\hyphenation{cal-or-i-me-ter}
\hyphenation{de-vices}
\RCS$Revision: 447187 $
\RCS$HeadURL: svn+ssh://svn.cern.ch/reps/tdr2/papers/SUS-17-001/trunk/SUS-17-001.tex $
\RCS$Id: SUS-17-001.tex 447187 2018-02-21 13:12:15Z schoef $
\newlength\cmsFigWidth
\ifthenelse{\boolean{cms@external}}{\setlength\cmsFigWidth{0.98\columnwidth}}{\setlength\cmsFigWidth{0.8\textwidth}}
\ifthenelse{\boolean{cms@external}}{\providecommand{\cmsLeft}{upper\xspace}}{\providecommand{\cmsLeft}{left\xspace}}
\ifthenelse{\boolean{cms@external}}{\providecommand{\cmsRight}{lower\xspace}}{\providecommand{\cmsRight}{right\xspace}}
\ifthenelse{\boolean{cms@external}}{\providecommand{\cmsLLeft}{Upper\xspace}}{\providecommand{\cmsLLeft}{Left\xspace}}
\ifthenelse{\boolean{cms@external}}{\providecommand{\cmsRRight}{Lower\xspace}}{\providecommand{\cmsRRight}{Right\xspace}}
\ifthenelse{\boolean{cms@external}}{\providecommand{\CL}{C.L.\xspace}}{\providecommand{\CL}{CL\xspace}}
\newlength\cmsTabSkip\setlength\cmsTabSkip{3pt}
\ifthenelse{\boolean{cms@external}}{\providecommand{\cmsTableResize}[1]{#1}}{\providecommand{\cmsTableResize}[1]{\resizebox{\textwidth}{!}{#1}}}

\newcommand{\Njets}{\ensuremath{N_\text{jets}}\xspace}
\newcommand{\Nbjets}{\ensuremath{N_\text{\PQb jets}}\xspace}
\newcommand{\Ttt}{\ensuremath{\text{T2}\PQt\PQt}\xspace}
\newcommand{\TbW}{\ensuremath{\text{T2}{\PQb\PW}}\xspace}
\newcommand{\Tbbllnunu}{\ensuremath{\text{T8}\PQb\PQb\ell\ell\PGn\PGn}\xspace}
\newcommand{\ttZ}{\ensuremath{\ttbar\cPZ}\xspace}
\newcommand{\tqZ}{\ensuremath{\PQt\PQq\cPZ}\xspace}
\newcommand{\ttG}{\ensuremath{\ttbar\gamma}\xspace}
\newcommand{\ttW}{\ensuremath{\ttbar\PW}\xspace}
\newcommand{\ttH}{\ensuremath{\ttbar\PH}\xspace}
\newcommand{\ttX}{\ensuremath{\ttbar X}\xspace}
\newcommand{\pp}{\ensuremath{\Pp\Pp}\xspace}
\newcommand{\ptmissSig}{\ensuremath{S}\xspace}
\newcommand{\mt}{\ensuremath{M_{\text{T}}}\xspace}
\newcommand{\mtll}{\ensuremath{M_{\text{T2}}(\ell \ell)}\xspace}
\newcommand{\mtlblb}{\ensuremath{M_{\text{T2}}(\PQb\ell \PQb\ell )}\xspace}
\newcommand{\lumiGolden}{35.9\fbinv}

\cmsNoteHeader{SUS-17-001}
\title{Search for top squarks and dark matter particles in opposite-charge dilepton final states at \texorpdfstring{$\sqrt{s}= 13 \TeV$}{13 TeV}}

\date{\today}

\abstract{
A search for new physics is presented in final states with two oppositely charged leptons (electrons or muons), jets identified as originating from \PQb quarks, and missing transverse momentum (\ptmiss). The search uses proton-proton collision data at $\sqrt{s}=13\TeV$
amounting to 35.9\fbinv of integrated luminosity collected using the CMS detector in 2016.
Hypothetical signal events are efficiently separated from the dominant \ttbar background with requirements on \ptmiss and transverse mass variables.
No significant deviation is observed from the expected background.
Exclusion limits are set in the context of simplified supersymmetric models with pair-produced top squarks.
For top squarks, decaying exclusively to a top quark and a neutralino, exclusion limits are placed at 95\% confidence level on the mass of the lightest top squark up to 800\GeV and on the lightest neutralino up to 360\GeV.
These results, combined with searches in the single-lepton and all-jet final states, raise the exclusion limits up to 1050\GeV for the lightest top squark and up to 500\GeV for the lightest neutralino.
For top squarks undergoing a cascade decay through charginos and sleptons, the mass limits reach up to 1300\GeV for top squarks and up to 800\GeV for the lightest neutralino.
The results are also interpreted in a simplified model with a dark matter (DM) particle coupled to the top quark through a scalar or pseudoscalar mediator.
For light DM, mediator masses up to 100\,(50)\GeV are excluded for scalar (pseudoscalar) mediators.
The result for the scalar mediator achieves some of the most stringent limits to date in this model.
}

\hypersetup{
pdfauthor={CMS Collaboration},
pdftitle={Search for new physics in the context of top quarks in the dilepton final state at sqrt(s)=13TeV},
pdfsubject={CMS},
pdfkeywords={CMS, physics, supersymmetry, top quarks}}

\maketitle

\section{Introduction}
\label{sec:Introduction}
The top quark couples to the Higgs boson more strongly than other fermions because of its large mass.
As a result, it plays a prominent role in the so-called hierarchy problem~\cite{Witten:1981nf,Dimopoulos:1981zb} of the standard model (SM) of particle physics, since its dominant contribution in the loop corrections to the Higgs boson mass exposes the theory to higher energy scales present in nature.
Supersymmetry (SUSY)~\cite{Ramond:1971gb,Golfand:1971iw,Neveu:1971rx,Volkov:1972jx,Wess:1973kz,Wess:1974tw,Fayet:1974pd,Nilles:1983ge} is a well-motivated theory
beyond the SM that provides a solution to the hierarchy problem. In addition, in $R$-parity conserving SUSY~\cite{FARRAR1978575}, the
lightest SUSY particle (LSP) is stable and can be a viable dark matter (DM) candidate, assuming it is neutral and weakly interacting.
Presently, the lighter SUSY particles may have masses in the \TeV range and therefore could be produced in proton-proton (\pp) collisions at the CERN LHC.
The scalar partners of the right- and left-handed top quarks, the top squarks $\PSQt_{R}$ and $\PSQt_{L}$, can be among these particles.
These two states mix into the mass eigenstates $\PSQtDo$ and $\PSQtDt$.
The lighter one, $\PSQtDo$, could be within the LHC energy reach to provide a natural solution to the hierarchy problem~\cite{Papucci:2011wy}, which strongly motivates searches for top squark production.

In this paper, we present a search for top squark pair production in a final state with two leptons (electrons or muons),
hadronic jets identified as originating from \PQb quarks, and significant transverse momentum imbalance.
The search is performed using data from \pp collisions collected with the CMS detector at the LHC during 2016 at a center-of-mass energy of 13\TeV, corresponding to an integrated luminosity of \lumiGolden.
We employ an efficient background reduction strategy that suppresses the large background from SM $\ttbar$ events by several orders of magnitude through use of dedicated transverse-mass variables~\cite{Smith:1983aa,Lester:1999tx}.
The predicted SM backgrounds in the various search regions are validated in data control samples orthogonal in selection to the signal regions in data.

\begin{figure*}[htb]
\centering
\includegraphics[width=0.4\textwidth]{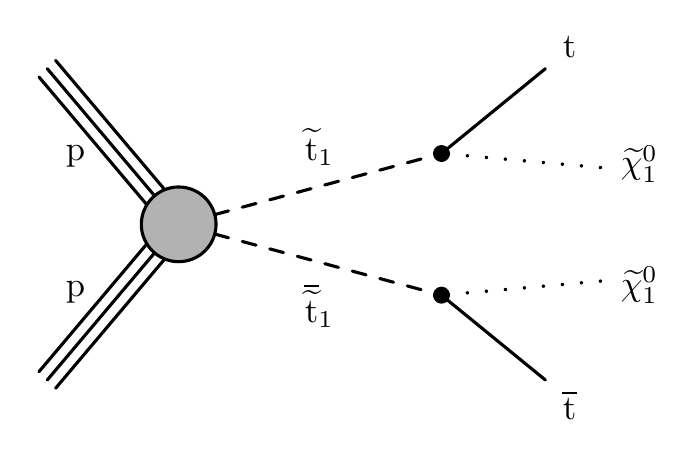}
\includegraphics[width=0.4\textwidth]{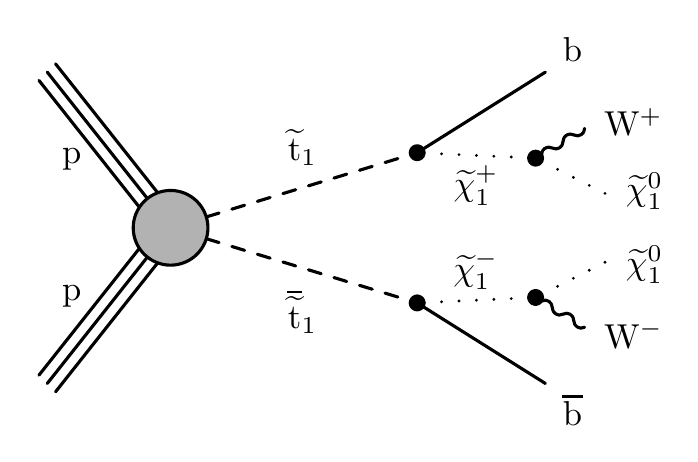}
\includegraphics[width=0.4\textwidth]{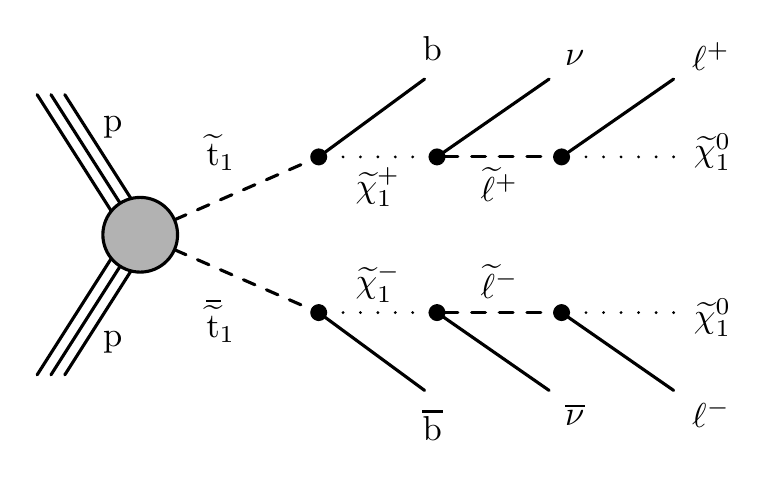}
\includegraphics[width=0.45\textwidth]{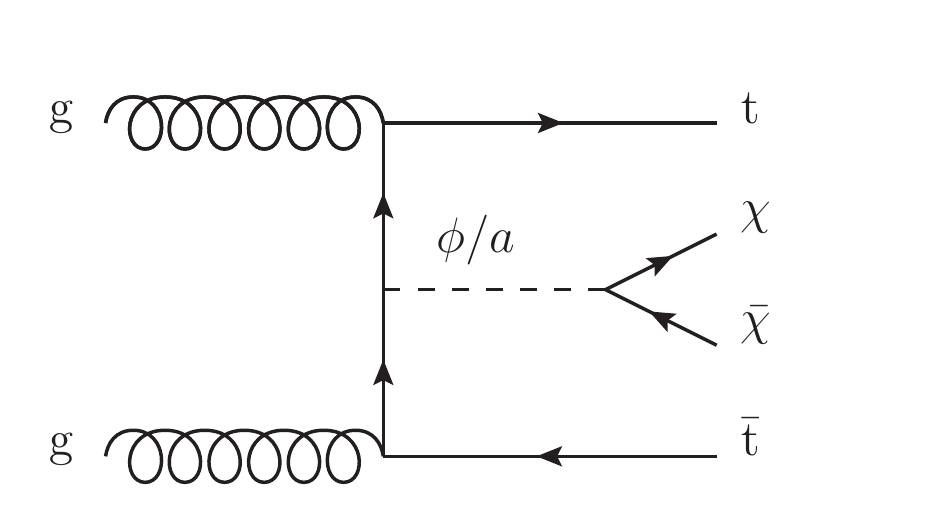}
\caption{Diagrams for simplified SUSY models and for direct DM production: strong production of top squark pairs $\PSQtDo\PASQt_{1}$,
where each top squark decays to a top quark and a \PSGczDo (\Ttt model, upper left),
or where each top squark decays into a \PQb quark and an intermediate \PSGcpmDo that further decays into a \PW{} boson and a \PSGczDo (\TbW model, upper right),
or to a neutrino and an intermediate slepton $\PGn\widetilde{\ell}^{\pm}$ that yield $\PGn\PSGczDo$ and an $\ell^\pm$ from the virtual slepton decay (\Tbbllnunu model, lower left).
Direct DM production through scalar or pseudoscalar mediators in association with top quarks is shown at the lower right.
}
\label{fig:fey}
\end{figure*}

The search is interpreted in simplified models~\cite{Alwall:2008ag,Alwall:2008va,Alves:2011wf} describing the strong production of pairs of top squarks.
We consider different decay modes, following the naming convention in Ref.~\cite{Chatrchyan:2013sza}.
In the \Ttt model (Fig.~\ref{fig:fey}, upper left), each top squark decays into a top quark and the lightest neutralino \PSGczDo, which is the LSP.
Alternatively, we consider the \TbW model (Fig.~\ref{fig:fey}, upper right), where both top squarks decay into a \PQb quark and an intermediate chargino (\PSGcpmDo) which further decays into a \PW~boson and an LSP.
In both models, leptonic decays of the two $\text{W}$ bosons provide a low-background final state with two oppositely charged leptons,
jets from \PQb quarks, and significant transverse momentum imbalance due to undetected LSPs and neutrinos.
The obtained results are then combined with results from searches in the same dataset in the single-lepton and all-jet final states~\cite{Sirunyan:2017xse,stop0L2016}.
Finally, we consider for the first time the \Tbbllnunu model (Fig.~\ref{fig:fey}, lower left), where both top squarks decay via charginos to sleptons and, subsequently, to neutralinos leading to a final state with the same particle content as in the \Ttt model.
Here, sleptons are the SUSY partners of leptons, and the branching fraction of the chargino is taken to be identical for all three flavors.
In this way, and contrary to the \Ttt and \TbW models, the branching fraction to a pair of oppositely charged leptons is 100\% when decays to $\tau$ leptons are included.
Searches based on \Ttt and \TbW models using 8 and 13\TeV \pp{} collision data recorded before 2016 were published
by the CMS~\cite{Chatrchyan:2013xna,Khachatryan:2016pup,Sirunyan:2016jpr} and the ATLAS~\cite{Aaboud:2017nfd,Aad:2015pfx,Aad:2014kra,Aad:2014qaa,Aaboud:2016lwz} experiments, with a $\PSQtDo$ mass excluded up to 700\GeV in the \Ttt model.

As an alternative to the SUSY hypothesis, we also interpret the search in a simplified model where a DM candidate $\PGc$ interacts with SM particles through a scalar ($\phi$) or pseudoscalar (\Pa) mediator~\cite{Lin:2013sca,Buckley:2014fba,Haisch:2015ioa,Arina:2016cqj,Abercrombie:2015wmb}.
Assuming minimal flavor violation~\cite{D'Ambrosio:2002ex,Isidori:2012ts}, the DM particles are dominantly produced in pairs in association with a \ttbar pair (Fig.~\ref{fig:fey}, lower right).
This model predicts therefore the same final state as considered in SUSY phenomenology, with the transverse momentum imbalance provided by the DM particles.
Prior searches for such direct DM production via scalar and pseudoscalar mediators have been carried out at the LHC with 8\TeV data~\cite{Aad:2014vea,Khachatryan:2015nua}, and more recently with 13\TeV data~\cite{Sirunyan:2017xgm,Sirunyan:2017hci,Aaboud:2017rzf}.

\section{The CMS detector}
\label{sec:CMS}
The central feature of the CMS detector is a
superconducting solenoid of 6\unit{m} internal diameter, providing a
magnetic field of 3.8\unit{T}. A silicon pixel and a silicon strip tracker, a
lead tungstate crystal electromagnetic calorimeter (ECAL), and a
brass and scintillator hadron calorimeter, each comprising a barrel and
two end sections reside within the solenoid volume. Muons are measured in gas-ionization detectors
embedded in the magnet steel flux-return yoke outside the
solenoid.
Extensive forward calorimetry complements the coverage
provided by the barrel and endcap detectors that improve the measurement of the imbalance in transverse momentum.
A more detailed description of the CMS detector, together with a definition
of the coordinate system and the kinematic variables,
can be found in Ref.~\cite{Chatrchyan:2008zzk}.

\section {Event samples}
\label{sec:samples}

During data taking, events are selected for offline analysis by different trigger algorithms that require the presence of one or two leptons (electrons or muons).
For the dilepton triggers, which accept the majority of events with two leptons, the thresholds are 23\GeV on the leading lepton \pt and 12\GeV (electron) or 8\GeV (muon) on the subleading lepton \pt.
Efficiencies of the dilepton triggers are measured in data events that are selected independently of the leptons, based on the presence of jets and requirements on the transverse momentum imbalance (\ptmiss).
Typical values range from 95 to 99\%, depending on the momenta and pseudorapidities ($\eta$) of the two leptons and
are applied as scale factors to simulated events.

The top quark antiquark pair production (\ttbar) and $t$-channel single top quark background samples are simulated using the \POWHEG v2~\cite{powheg1,powheg2} event generator, and are normalized to next-to-next-to-leading order
(NNLO) cross sections~\cite{Czakon:2011xx,Aliev:2010zk,Beneke:2011mq,Czakon:2012zr,Czakon:2012pz,Czakon:2013goa,Kant:2014oha,Czakon:2011xx}.
Events with single top quarks produced in association with \PW{} bosons ($\PQt\PW$) are simulated using \POWHEG v1~\cite{Re:2010bp} and normalized to the NNLO cross section.
Drell--Yan and \ttZ events are generated with \MGvATNLO v2.2.2~\cite{Alwall:2014hca} at leading order (LO) and next-to-leading order (NLO), respectively, and their cross sections are computed at NNLO~\cite{Gavin:2010az} and NLO~\cite{Garzelli:2012bn}, respectively.
The processes \ttW, $\PQt\Z\PQq$, $\ttbar\gamma$, and the triboson processes are generated using \MGvATNLO at NLO, while $\PQt\PW\Z$ is generated at LO.
The diboson and \ttH processes are generated using \POWHEG v2 at NLO.
These processes are normalized to the most precise available cross section, corresponding to NLO accuracy in most cases.

Generated events are interfaced with \PYTHIA v8.205~\cite{Sjostrand:2014zea} using the CUETP8M1 tune~\cite{Skands:2014pea,CMS-PAS-GEN-14-001} or, for \ttbar and \ttH backgrounds, the CUETP8M2 tune, to simulate parton showering, hadronization, and  the underlying event.
The NNPDF3.0~\cite{Ball:2014uwa} parton distribution functions (PDFs) at NLO and LO are used consistently with NLO and LO event generators, respectively.
The events are subsequently processed with a \GEANTfour-based simulation model~\cite{Geant} of the CMS detector.

Signal samples including top squark pairs are generated with \MGvATNLO at LO precision, interfaced with \PYTHIA.
For the \Ttt and \TbW models, the top squark mass is varied from 150 to 1200\GeV and the mass of the LSP is scanned from 1 to 650\GeV.
The mass of the chargino in the \TbW model is assumed to be equal to the mean of the masses of the top squark and the lightest neutralino.
For the \Tbbllnunu model, we vary the top squark mass between 200 to 1400\GeV and the mass of the LSP from 1 to 1000\GeV.
The masses of the intermediate chargino and slepton states in the \Tbbllnunu model are chosen as follows:
for the chargino mass we assume $m_{\PSGcpDo} = (m_{\PSQtDo} + m_{\PSGczDo})/2$, while the
slepton masses are chosen by the three values $x = 0.95$, 0.50, 0.05 in $m_{\tilde{\ell}} = x \, (m_{\PSGcpDo} - m_{\PSGczDo}) + m_{\PSGczDo}$.
The signal production cross sections are normalized to NLO plus next-to-leading logarithmic (NLL) accuracy~\cite{Borschensky:2014cia}.
Simulation of the detector response is performed using the CMS fast detector simulation~\cite{fastsim}.

For the simplified model of ${\rm t\bar{t}}$+DM production, \MGvATNLO is used at LO to generate events with at most one additional parton from initial-state radiation.
We follow the recommendations from Ref.~\cite{Abercrombie:2015wmb}:
the DM particle is taken to be a Dirac fermion, while the spin-0 mediator can have either scalar or pseudoscalar couplings to both quarks and DM, ignoring mixing with the SM Higgs boson in the scalar case.
Yukawa couplings proportional to $g_{\PQq} m_{\PQq}$ are assumed between the mediator and the quarks of mass $m_{\PQq}$, where the coupling strength $g_{\PQq}$ is taken to be 1 and assumed to be flavor universal.
The coupling strength $g_\mathrm{DM}$ of the mediator to the DM particles is also set to 1.
The aforementioned \GEANTfour-based detector simulation is used for this signal.

All simulated samples include the simulation of so-called pileup from the presence of additional \pp collisions in simultaneous or preceding bunch crossings,
 and are reweighted according to the distribution of the true number of interactions in the main collision's bunch crossing.

\section{Object selection}
\label{sec:objects}
Offline event reconstruction uses the CMS particle-flow (PF) algorithm~\cite{Sirunyan:2017ulk},
yielding a consistent set of electron~\cite{Khachatryan:2015hwa}, muon~\cite{Chatrchyan:2012xi}, charged and neutral hadron, and photon candidates.
These particles are defined with respect to the primary $\Pp\Pp$ interaction vertex, chosen to have the largest value of summed physics-object $\pt^2$, where these physics objects are reconstructed by a jet finding algorithm~\cite{Cacciari:2008gp,Cacciari:2011ma} applied to all charged tracks associated with the vertex.

Electron candidates are reconstructed using tracking and ECAL information, by combining the clusters of energy deposits in the ECAL with Gaussian sum filter tracks~\cite{Khachatryan:2015hwa}.
The electron identification is performed using shower shape variables, track-cluster matching variables, and track quality
variables. The selection is optimized to identify electrons from the decay of SM bosons with a 70\% efficiency while rejecting electron candidates originating from jets.
To reject electrons originating from photon conversion inside the detector, electrons are required to have all possible hits in the innermost tracker layers and to be incompatible with any conversion-like secondary vertices.
Identification of muon candidates is performed using the quality of the geometrical matching between the measurements of the
tracker and the muon system~\cite{Chatrchyan:2012xi}.

All lepton candidates are required to satisfy $\pt > 25 (20) \GeV$ for the leading (subleading) lepton and $\abs{\eta} < 2.4$.
Consistency of the lepton track with the selected primary vertex is enforced by vetoing lepton candidates
whose tracks have a significance of the transverse impact parameter above 4. Here, the impact parameter is the minimum three-dimensional distance between the lepton trajectory and the primary vertex.
Its significance is defined as the ratio of the impact parameter to its uncertainty.
The longitudinal displacement from the primary collision vertex must also be less than $0.1 \cm$.

Lepton candidates are required to be isolated. For each candidate a cone with radius $\Delta R=\sqrt{\smash[b]{(\Delta\eta)}^2+\smash[b]{(\Delta\phi)}^2}=0.3$ (where $\phi$ is azimuthal angle in radians)
around the track direction at the event vertex is constructed.
The relative isolation ($I_{\textrm{rel, 0.3}}$) is defined as the scalar \pt sum, normalized to the lepton \pt, of photons and neutral and charged hadrons reconstructed by the PF algorithm within this cone.
In order to reduce dependence on the number of pileup interactions,
charged hadron candidates are included in the sum only if they are consistent with originating from the selected primary vertex in the event. The contribution of neutral particles from pileup events is estimated
following the method described in Ref.~\cite{Khachatryan:2015hwa}, and subtracted from the isolation sum.
For a lepton candidate to be isolated, $I_{\textrm{rel, 0.3}}$ has to be smaller than 0.12.

Jets are clustered from PF candidates using the anti-\kt algorithm~\cite{Cacciari:2008gp} with a distance parameter of $R=0.4$. The influence of pileup is mitigated using the charged hadron subtraction technique, by
subtracting the energy of charged hadrons associated to vertices other than the primary vertex.
Jet momenta are then further calibrated, accounting for deposits from neutral pileup particles and the imperfect detector response~\cite{Khachatryan:2016kdb}, and quality criteria are applied for jets with $\pt> 30\GeV$ and $\abs{\eta}<2.4$.
To arbitrate between jets and leptons, jets that are found within a cone with radius $\Delta R=0.4$ around any isolated lepton are removed from the set of selected jets.
The scalar \pt sum of the jets that pass this selection is denoted by $\HT$.

The vector \ptvecmiss is defined as the negative vector \pt sum of all PF candidates reconstructed in an event and is corrected to account for the jet energy corrections. Its magnitude is denoted by \ptmiss.
Events with possible contributions from beam halo processes or anomalous noise in the calorimeter are rejected using dedicated filters~\cite{Khachatryan:2014gga}.

A multivariate \PQb tagging discriminator CSVv2~\cite{CMS:2016kkf} is used to identify jets that originate from a \PQb quark (b jets).
The chosen ``medium'' working point has a mistag rate of approximately~1\% for light flavor jets and a corresponding \PQb tagging efficiency of 55\% to 65\% depending on jet transverse momentum and pseudorapidity~\cite{CMS:2016kkf}.

Scale factors are applied in simulation to take into account the differences of lepton reconstruction, identification and isolation as well as \PQb tagging efficiencies in data and simulation.
Typical corrections are less than 1\% per lepton and less than 10\% per b-tagged jet.

\section {Search strategy}
\label{sec:strategy}
We select events containing a pair of leptons with opposite charge, and we require the invariant mass of the lepton pair to be greater than 20\GeV, to suppress backgrounds with misidentified or nonprompt leptons from the hadronization of (heavy flavor) jets in multijet events.
Events with additional leptons with $\pt > 15 \GeV$ and
satisfying a looser isolation criterion of $I_{\text{rel, 0.3}}<0.4$ are vetoed.
In case of a same-flavor~(SF) lepton pair, we suppress contributions from SM Drell--Yan production with a requirement on the dilepton mass, $\abs{m_{\PZ} - m(\ell\ell)} > 15\GeV$, where $m(\ell\ell)$ is the invariant mass of the dilepton system and $m_{\PZ}$ is the mass of the \PZ boson.
To further suppress this and other vector boson backgrounds, we require the number of jets (\Njets) to be at least two and, among them, the number of \PQb jets (\Nbjets) to be at least one.
After additionally requiring $\ptmiss > 80\GeV$, a small background remains from events with vector bosons and highly energetic jets that are severely mismeasured.
We further reduce this background by defining $S=\ptmiss/\sqrt{\HT}$ and requiring $S>5\GeV^{1/2}$ and, furthermore, by placing a requirement on the angular separation of \ptvecmiss and the momenta
of the leading ($j_1$) and subleading ($j_2$) jets in the azimuthal plane.
The selection above is summarized in Table~\ref{Tab:baselineSel} and defines the event sample, which is dominated by events with top quark pairs that decay to a dilepton final state.

\begin{table}[htb]
\centering
\topcaption{Overview of the preselection requirements.}
\label{Tab:baselineSel}
\begin{scotch}{ll}
Leptons & = 2 (\Pe{} or $\mu$), oppositely charged  \\
$m(\ell\ell)$ & $>$20\GeV\\
$\abs{m_{\Z} - m(\ell\ell)}$ & $>$15\GeV, same flavor only\\
$\Njets$ & $\geq$2\\
$\Nbjets$ & $\geq$1\\
$\ptmiss$ & $>$80\GeV  \\
$S$&$>$5\GeV${}^{1/2}$   \\
$\cos\Delta\phi(\ptmiss, j_{1})$ & $<$0.80 \\
$\cos\Delta\phi(\ptmiss, j_{2})$ & $<$0.96 \\
\end{scotch}
\end{table}

The main search variable in this analysis is
\ifthenelse{\boolean{cms@external}}{
 \begin{multline}
\mtll = \min_{\ptvecmiss{}^{1} + \ptvecmiss{}^{2} = \ptvecmiss} \Bigl( \max \bigl[ \mt(\ptvec^{\text{vis}1},\\
\ptvecmiss{}^{1}), \mt(\ptvec^{\text{vis}2},\ptvecmiss{}^{2}) \bigr] \Bigr),
\label{eq:MT2def}
\end{multline}
}{
 \begin{equation}
\mtll = \min_{\ptvecmiss{}^{1} + \ptvecmiss{}^{2} = \ptvecmiss} \left( \max \left[ \mt(\ptvec^{\text{vis}1},\ptvecmiss{}^{1}), \mt(\ptvec^{\text{vis}2},\ptvecmiss{}^{2}) \right] \right),
\label{eq:MT2def}
\end{equation}
}
where the choice $\ptvec^{\text{vis}1,2}=\ptvec^{\ell1,2}$ corresponds to the definition introduced in Ref.~\cite{Burns:2008va} and used in Ref.~\cite{Khachatryan:2016pup}.
The calculation of \mtll is performed through the algorithm discussed in Ref.~\cite{Cheng:2008hk} assuming vanishing mass for the undetected particles.
Under the hypothesis of a well-reconstructed dileptonic \ttbar or $\PW\PW$ event, the minimization in Eq.~\ref{eq:MT2def} encompasses the correct neutrino momenta, and thus \mtll has an endpoint at the parent particle's mass~\cite{Lester:1999tx}, here $m_\text{W}$.
When the azimuthal angle of \ptvecmiss falls within the smaller of the two opening angles defined by the leptons in the transverse plane,
it follows that \mtll vanishes because the minimization procedure will find a partitioning where $\ptvecmiss{}^{1,2}$ and $\ptvec^{\ell1,2}$ are both parallel.

\begin{figure*}[htbp]
\centering
\includegraphics[width=0.32\textwidth]{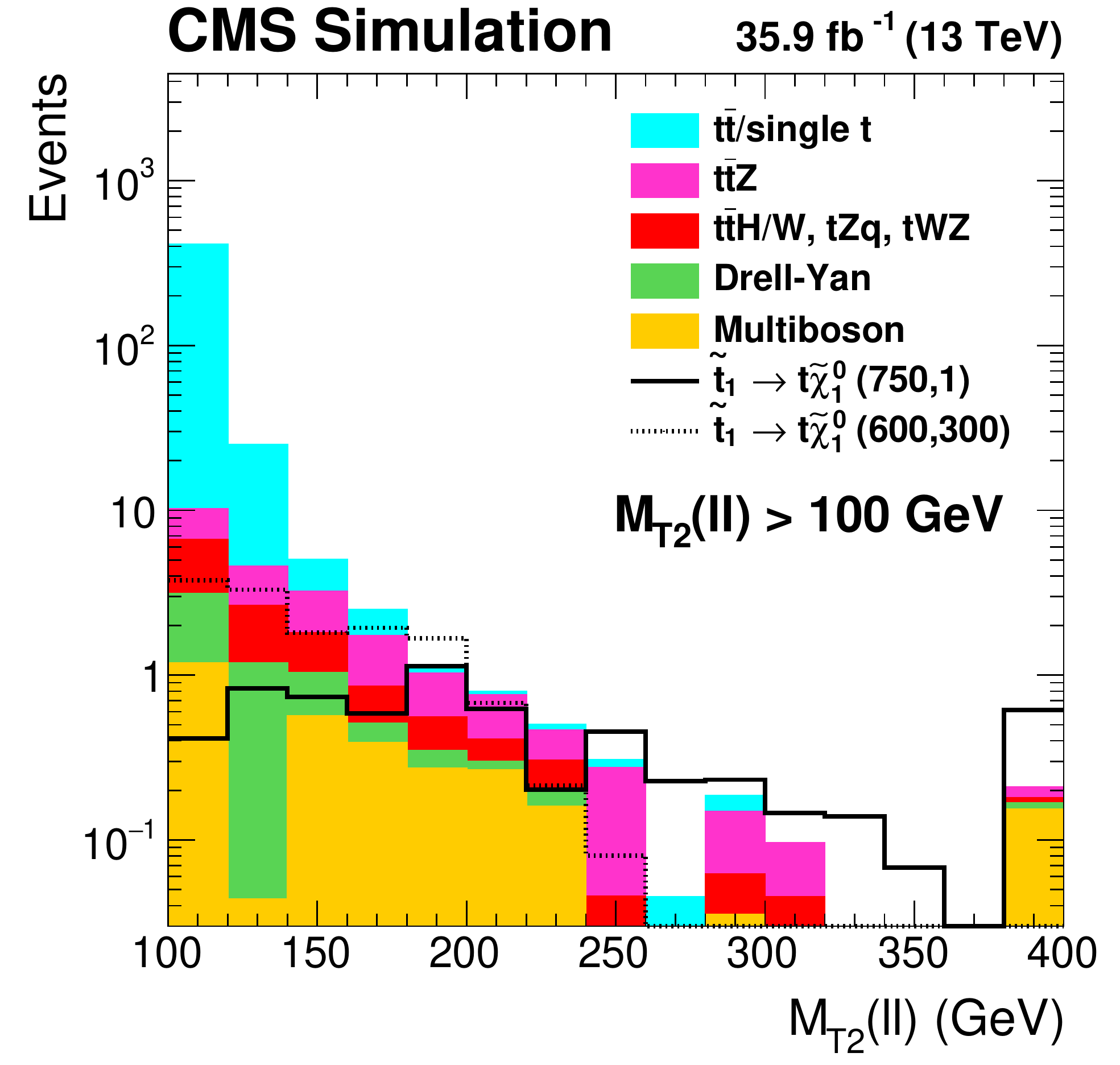}
\includegraphics[width=0.32\textwidth]{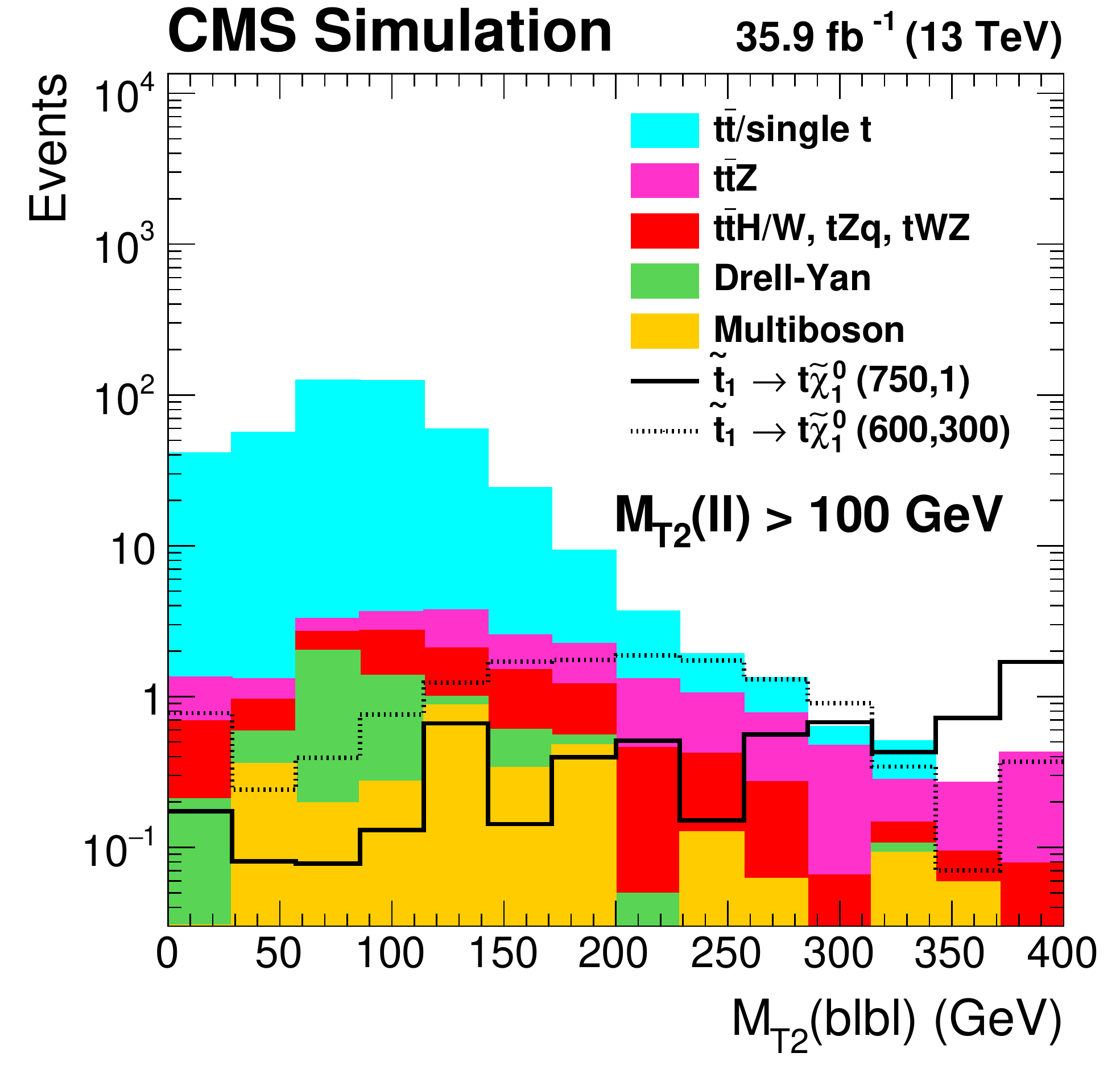}
\includegraphics[width=0.32\textwidth]{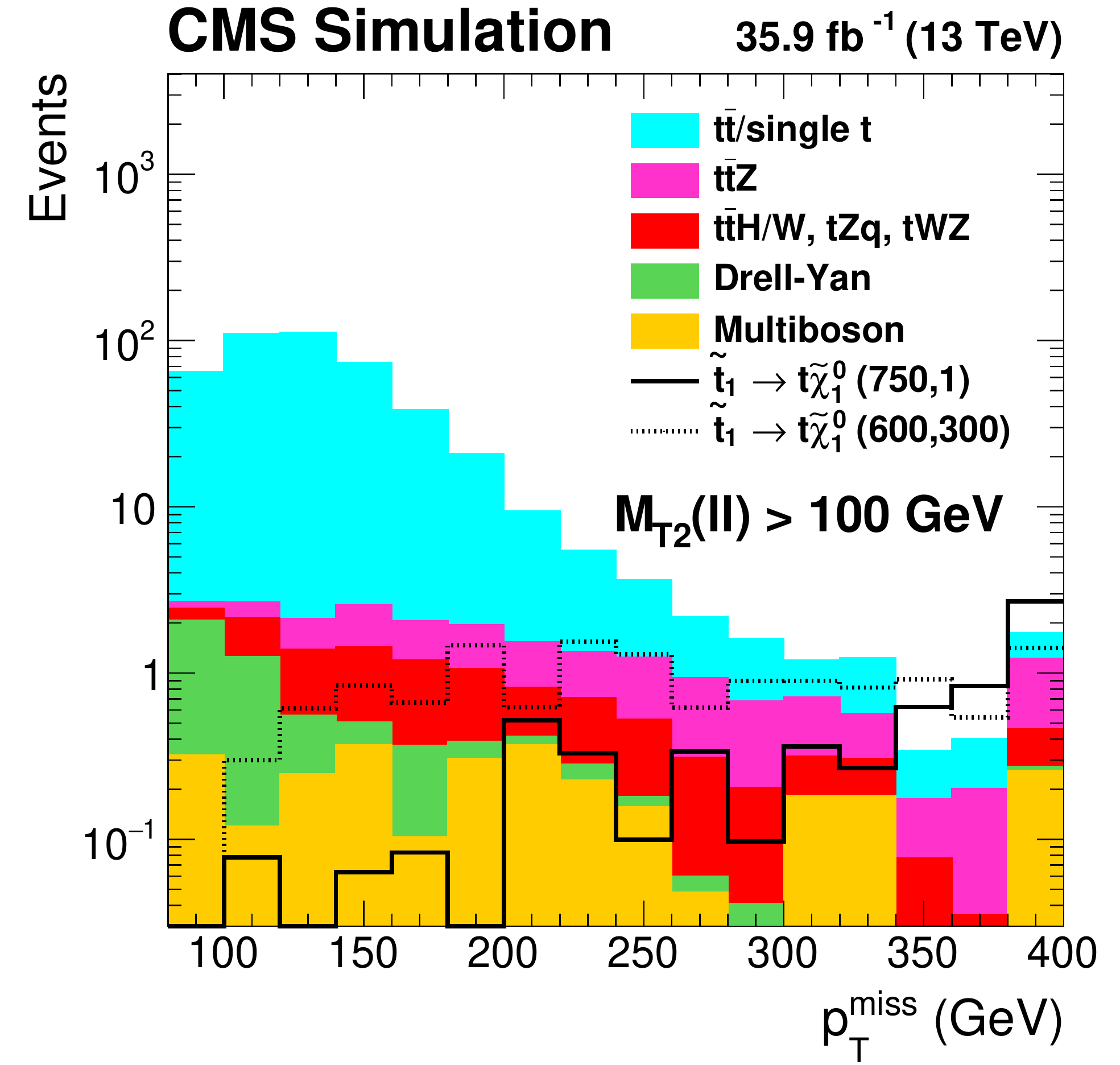}
\caption{Distributions of \mtll~(left), \mtlblb~(center), and \ptmiss~(right) in simulation after preselection and requiring $\mtll > 100\GeV$. A \Ttt signal is shown with masses $m_{\PSQt} = 750\GeV$ and $m_{\PSGczDo} = 1\GeV$, as well as a more compressed signal with $m_{\PSQt} = 600\GeV$ and $m_{\PSGczDo} = 300\GeV$.}
\label{fig:MT2llSimulation}
\end{figure*}

The key feature of this analysis is that the presence of additional invisible particles, \eg, the LSP $\PSGczDo$ or the DM particle $\PGc$, breaks the correlation between the \ptvecmiss and the lepton transverse momenta
that define the  \mtll endpoint. Hence, we expect the events predicted by the diagrams depicted in Fig.~\ref{fig:fey} to populate the tails of this distribution.
The distribution of \mtll in simulation after the preselection is shown in Fig.~\ref{fig:MT2llSimulation} (left) for $\mtll > 100\GeV$ and including a \Ttt signal with a mass configuration
with $m_{\PSQt} = 750\GeV$ and $m_{\PSGczDo} = 1\GeV$, as well as a more compressed signal scenario with $m_{\PSQt} = 600\GeV$ and $m_{\PSGczDo} = 300\GeV$.

We refine the analysis by using two more observables to define signal regions, \mtlblb and \ptmiss.
For \mtlblb, we choose \cite{Burns:2008va} $\ptvec^{\text{vis}1,2}=\ptvec^{\text{b} 1,2} + \ptvec^{\ell 1,2}$, which requires two b-tagged jets.
If only one \PQb tagged jet is found in the event, the jet with the highest \pt that does not pass the \PQb tagging selection is taken instead.
The ambiguity when pairing leptons with \PQb jets for \mtlblb is resolved by selecting the configuration which minimizes the maximum invariant mass of the two lepton-jet pairs.
Similar to the procedure to obtain \mtll, we break up \ptvecmiss into two parts and pair them with $\ptvec^{\text{vis}1,2}$ in order to define \mt, and then compute \mtlblb analogously to Eq. (\ref{eq:MT2def}).
For dileptonic \ttbar events, \mtlblb has an endpoint at the top quark mass.
After a tight threshold of $\mtll > 100\GeV$, both \mtlblb and \ptmiss still exhibit significant discrimination power.
This is shown in Fig.~\ref{fig:MT2llSimulation} (middle) for \mtlblb and Fig.~\ref{fig:MT2llSimulation} (right) for \ptmiss.

Based on sensitivity studies for a wide range of signal scenarios,
the signal regions listed in Table~\ref{table:srdefinition} are chosen.
These regions are further split depending on the flavor of the leptons into different- and same-flavor signal regions.
There is no overlap among the signal regions themselves or with background enriched regions (control regions) used in the following.

\begin{table*}[htb]
\centering
\topcaption{Definition of the signal regions. The regions are further split into different- and same-flavor regions.}
\label{table:srdefinition}
\cmsTableResize
{
\begin{scotch}{ccccc}
 $\mtlblb$ (\GeVns)                             & $\ptmiss$ (\GeVns)             	    	 & $100 < \mtll < 140\GeV$ 	      & $140 < \mtll < 240\GeV$         & $\mtll > 240\GeV$    	\\
 	\hline
\multirow{2}{*}{0--100}                & 80--200  &            SR0            &     SR6        & \multicolumn{1}{|c}{\multirow{7}{*}{SR12} }		              	\\	
                                                        & $>$200  &           SR1            &     SR7        &      \multicolumn{1}{|c}{}{}                                 					\\	[\cmsTabSkip]	
\multirow{2}{*}{100--200}              & 80--200   &          SR2             &     SR8        &          \multicolumn{1}{|c}{}{}                                					\\
                                                        & $>$200  &          SR3              &     SR9        &         \multicolumn{1}{|c}{}{}                                 					\\[\cmsTabSkip]
\multirow{2}{*}{ $>$200 }           & 80--200   &          SR4             &     SR10       &      \multicolumn{1}{|c}{}{}                                     					\\	
                                                        & $>$200   &         SR5              &    SR11       &     \multicolumn{1}{|c}{}{}                                      					\\
\end{scotch}
}
\end{table*}

\section {Background predictions}
\label{sec:backgrounds}
The major backgrounds from SM processes in the search regions after the event selection are single top quark and top quark pair events with either severely mismeasured \ptmiss or misidentified leptons.
Smaller contributions come from the same processes in association with a \PZ, \PW, or an \PH boson (\ttZ, \ttW, \ttH, \tqZ) and Drell--Yan and multiboson production ($\PW\PW$, $\PW\PZ$, $\PZ\PZ$, $\PW\PW\PW$, $\PW\PW\PZ$, $\PW\PZ\PZ$, and $\PZ\PZ\PZ$).
In the following, we discuss the estimation of these different background components.

\subsection{Top quark background}

\begin{figure}[htbp]
\centering
\includegraphics[width=0.45\textwidth]{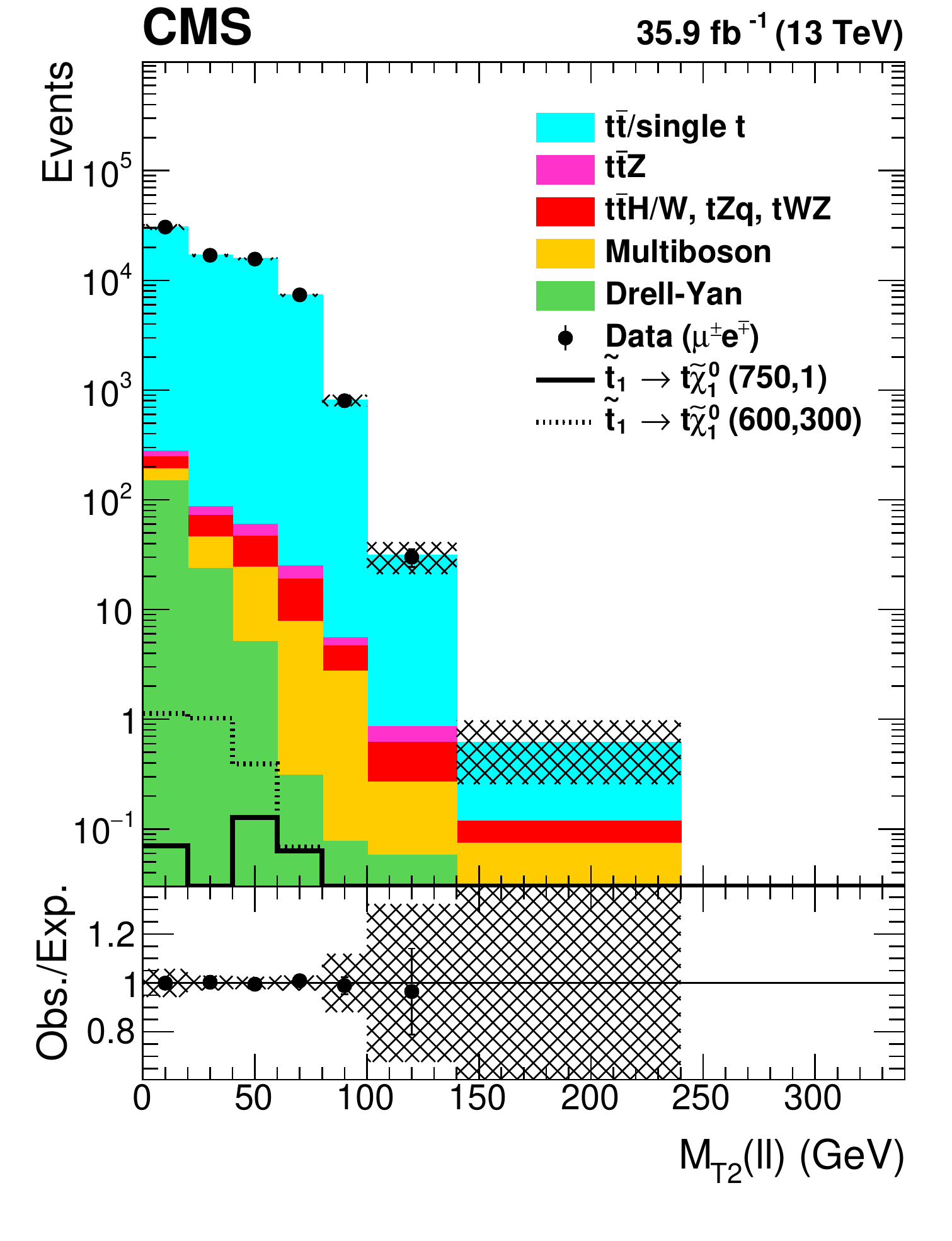}
\includegraphics[width=0.45\textwidth]{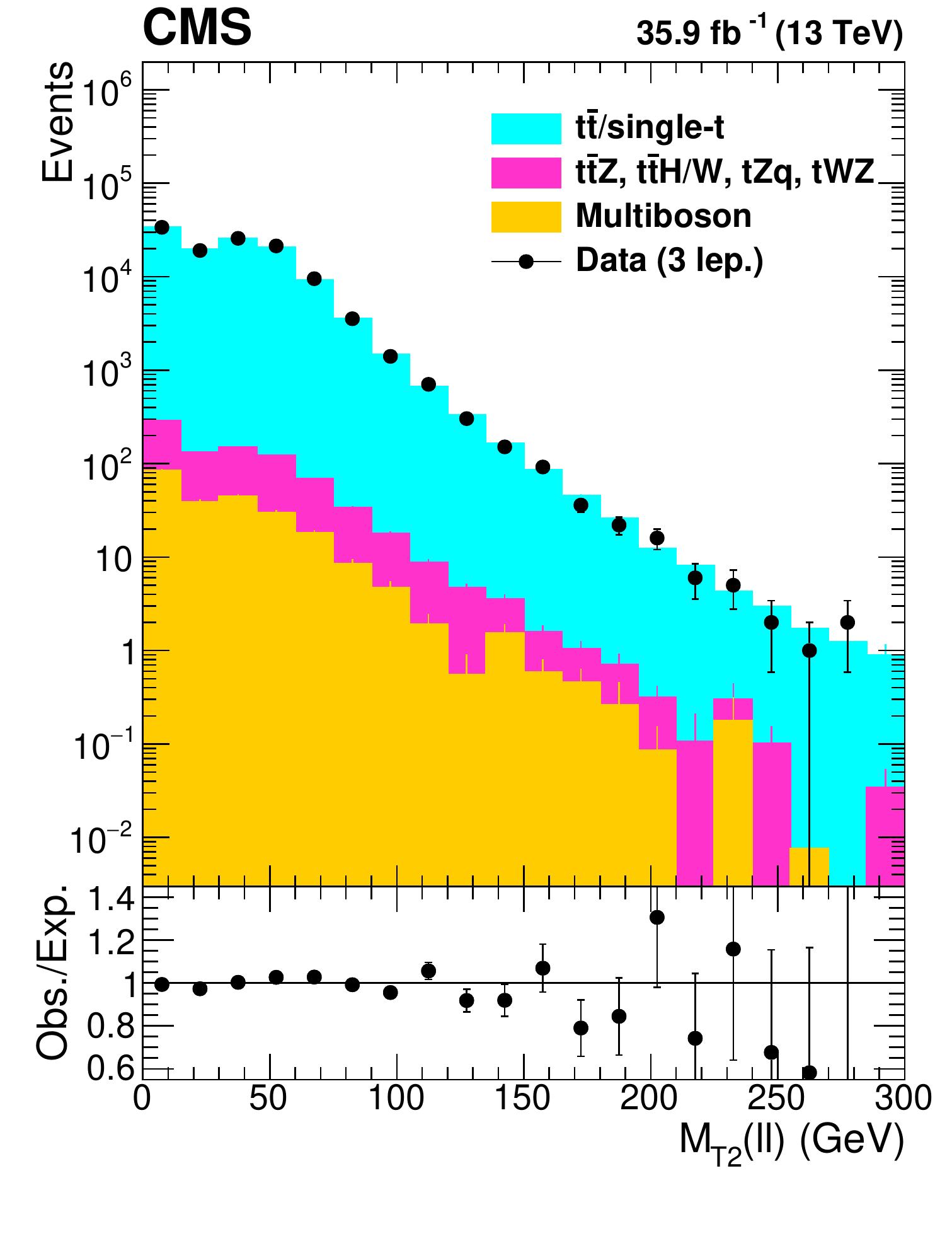}
\caption{\cmsLLeft: distribution of \mtll in a control region enriched in \ttbar events and defined by $\Njets \geq 2$, $\Nbjets \geq 1 $, and $\ptmiss < 80\GeV$. The hatched band shows the uncertainties from experimental
effects, as described in Section~\ref{sec:systematics}.
\cmsRRight: distribution of \mtll after swapping an isolated lepton with an additional non-isolated lepton, as described in the text.
For both plots, simulated yields are normalized to data using the yields in the $\mtll<100 \GeV$ region. }
\label{fig:ttBar_controlPlots}
\end{figure}

Events containing single or pair-produced top quarks populate low regions in the distributions of the three analysis variables \mtll, \mtlblb, and \ptmiss if the momenta in the events are well measured.
Studies based on simulation show two main sources of top quark background in the signal regions.
First, a severe mismeasurement of jet energy caused by the loss of photons and neutral hadrons showering in masked channels of the calorimeters
can induce large \ptmiss mismeasurement and promote an otherwise well-measured event to the signal regions.
Additionally, neutrinos with high \pt within jets cause mismeasurements of the jet \pt.
A control region requiring same-flavor leptons satisfying $\abs{m(\ell\ell)-m_{\PZ}}<15\GeV$ is used to constrain any mismodeling of this rare effect by comparing the \ptmiss tail between data and simulation.
It is found that the simulation predicts well such mismeasurements, and no sign of unaccounted effects in the \ptmiss measurement is observed.
Furthermore, the modeling of the tail of the analysis variable distributions is validated in control regions that invert the requirement on one or more of the following variables: \ptmiss with no requirement on \ptmissSig,
\Nbjets, and \Njets.
As an example, Fig.~\ref{fig:ttBar_controlPlots} (\cmsLeft) shows the \mtll distribution in the different-flavor channel with $\Nbjets\geq 1$, $\Njets\geq 2$, $\ptmiss<80\GeV$, and no requirement on \ptmissSig.
No significant sign of mismodeling is found in any of the control regions over at least three orders of magnitude in event yields.
The uncertainties from experimental effects, as described in Section~\ref{sec:systematics}, are shown with a hatched band.

Second, an electron or muon may fail the identification requirements, or the event may have a $\PGt$ lepton produced in a \PW~boson decay.
If there is a nonprompt lepton from the hadronization of a \PQb quark or a charged hadron misidentified as a lepton selected in the same event, the reconstructed value
for \mtll is not bound by the \PW{} mass.
To validate the modeling of this contribution, we select events with one additional lepton satisfying loose isolation requirements on top of the selection in Table~\ref{Tab:baselineSel}.
In order to mimic the lost prompt lepton background, we recompute \mtll  by combining each of the isolated leptons with the extra lepton in both data and simulation.
Since the transverse momentum balance is not significantly changed by lepton misidentification, the \ptmiss observable is not modified.
The resulting \mtll distribution is shown in Fig.~\ref{fig:ttBar_controlPlots} (\cmsRight) and serves as a validation of the modeling of the lost lepton background.
We observe overall good agreement between simulation and data, indicating that simulation describes such backgrounds well.

Top quark backgrounds are split into three categories in the signal regions and uncertainties related to them are assigned based on the agreement of data and simulation in the studies above.
The first category consists of events which are promoted to the \mtll tail due to Gaussian jet energy mismeasurements within approximately twice the jet energy resolution.
It comprises 25--55\% of the top quark background, depending on the signal region, and we assign a 15\% uncertainty in the yield of this fraction.
The second category, 40--50\% of the total top quark background yield, contains events with jets with more severe energy mismeasurements.
A 30\% uncertainty, based on studies in control regions, is assigned to the yield of events.
Events containing misidentified electrons or muons constitute 1--25\% of the top quark background, and based on studies on the modeling of the misidentification rate, a 50\% uncertainty is assigned.
Finally, we proceed to predict the background from single top and top quark pair production by normalizing simulated distributions to the number of events in a data region
defined by the selection in Table~\ref{Tab:baselineSel} and an additional requirement of $\mtll <100\GeV$. 
In this way, experimental uncertainties affecting the overall normalization are largely reduced.

\subsection{Top quark + X background}

\begin{figure}[htbp]
\centering
\includegraphics[width=0.49\textwidth]{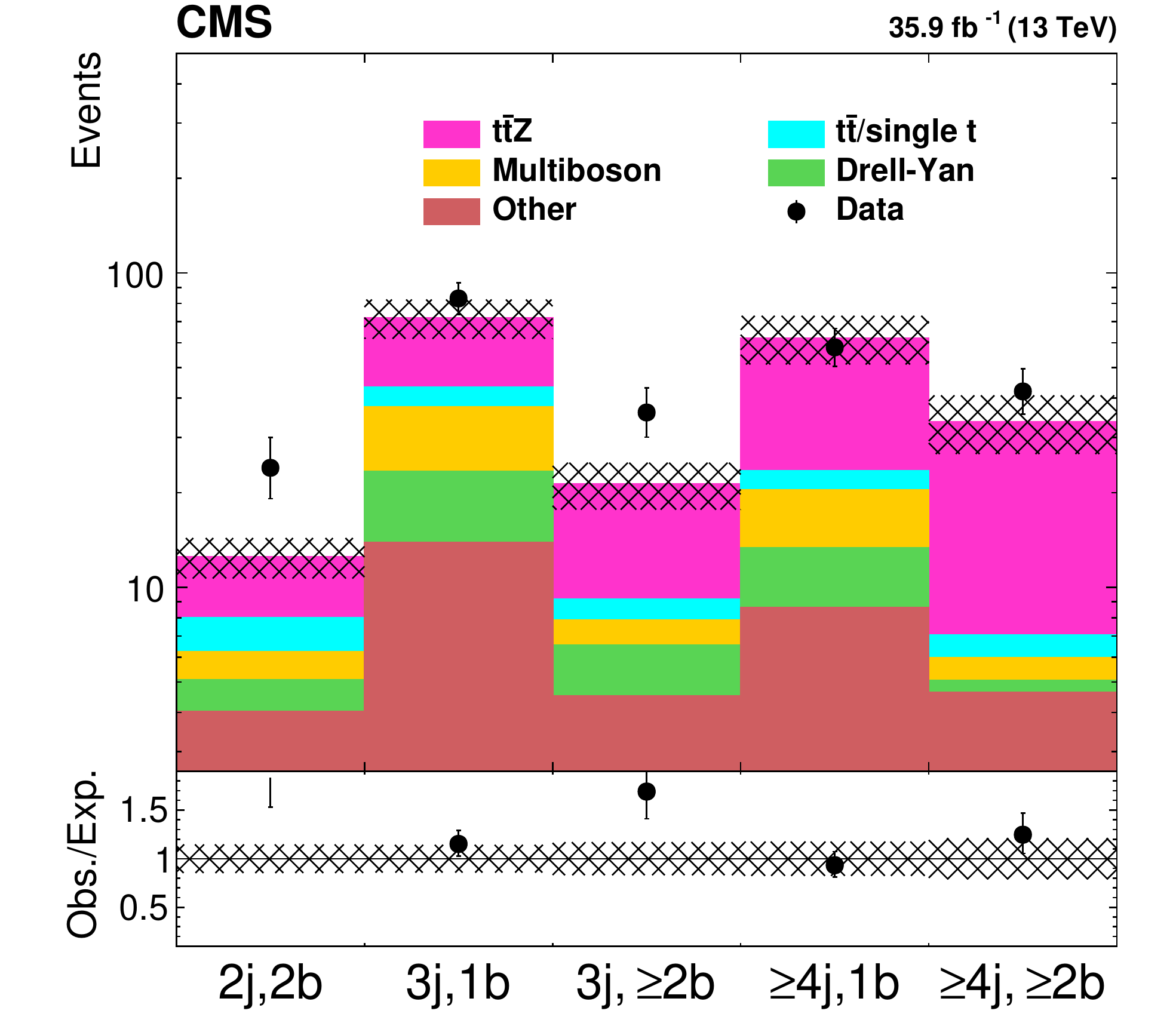}
\includegraphics[width=0.49\textwidth]{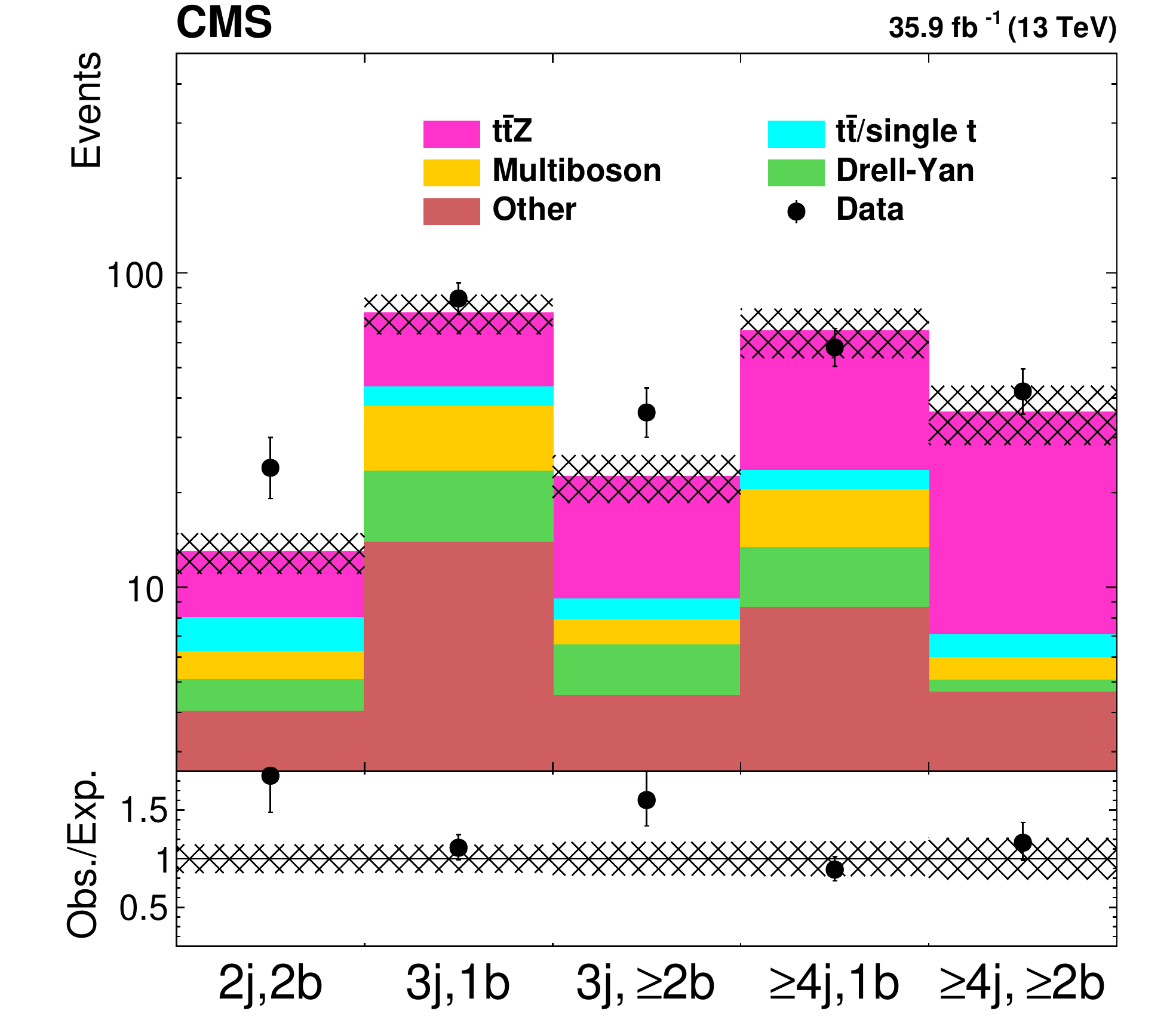}
\caption{Expected and observed yields in the five \ttZ control regions, which are defined by different requirements on the number of reconstructed jets and \PQb jets, before (\cmsLeft) and after the fit (\cmsRight).
The hatched band contains all uncertainties discussed in the text.}
\label{fig:3LttZ_controlPlots}
\end{figure}

Top quarks produced in association with a boson (\ttZ, \ttW, \ttH, \tqZ) form an irreducible background in decay channels where the bosons decay to leptons or neutrinos.
Among these, the \ttZ background, with $\Z\to \PGn \PAGn$ providing extra genuine \ptmiss, is the dominant one.
The overall normalization of this contribution is measured in the decay mode
\begin{equation*}
\ttZ \to  (\PQt\to\cPqb\ell^{\pm}\cPgn)(\PQt\to \PQb\cmsSymbolFace{jj})(\PZ \to \ell^{\pm}\ell ^{\mp})
\end{equation*}
in control regions with exactly three leptons ($\mu\mu\mu$, $\mu\mu\Pe$, $\mu\Pe\Pe$ and $\Pe\Pe\Pe$),
where the leading, subleading, and trailing lepton transverse momentum are required to satisfy thresholds of 40, 20, and 10\GeV, respectively.
All pairs of same-flavor leptons with opposite charge are required to satisfy $\abs{m(\ell\ell) - m_{\PZ} <10\GeV}$.
Five control regions requiring different \Njets and \Nbjets combinations are defined.
The simulated number of \ttZ events is fitted to the number of observed events in these regions.
The normalizations of other background components are allowed to vary within their uncertainties, and the values returned by the fit are consistent with the initial ones.
The number of events in the control regions in simulation and data is shown in Fig.~\ref{fig:3LttZ_controlPlots} before (left) and after (right) the fit.
Including systematic uncertainties, the fit yields a scale factor of $1.09 \pm 0.15$, which is then used to normalize the \ttZ background in the signal regions.
The scale factor uncertainty is fully accounted for in the background prediction.

Furthermore, we constrain a potential mismodeling of the \mtll and \ptmiss distributions for the \ttZ(with $\Z\to \PGn \PAGn$) background in a data control sample dominated by \ttG events, using the photon as a proxy for the \Z boson
and adding its momentum to the \ptmiss. To mitigate the difference between the massive \Z boson and the massless photon, the simulated photon momentum is reweighted to match the distribution of the \Z boson momentum.
After this procedure, we find good agreement between the simulated \ttG and \ttZ distributions.
Repeating the exercise on data, we find agreement within the statistical precision and assign a conservative additional uncertainty of 20\%.

\subsection{Drell--Yan and multiboson backgrounds}

Drell--Yan events constitute only a small background component after the analysis selection.
In order to measure the residual contribution, we select dilepton events where we invert the \Z boson veto, the \PQb jet requirements, and the angular separation requirements on jets and \ptvecmiss.
From simulation, this selection is expected to retain about 85\% Drell--Yan events, while the subleading contribution comes from multiboson events.
For each same-flavor signal region, we define a corresponding control region with the selections above and the signal region requirements on \mtll, \mtlblb, and \ptmiss.

\begin{figure*}[htbp]
\centering
\includegraphics[width=0.32\textwidth]{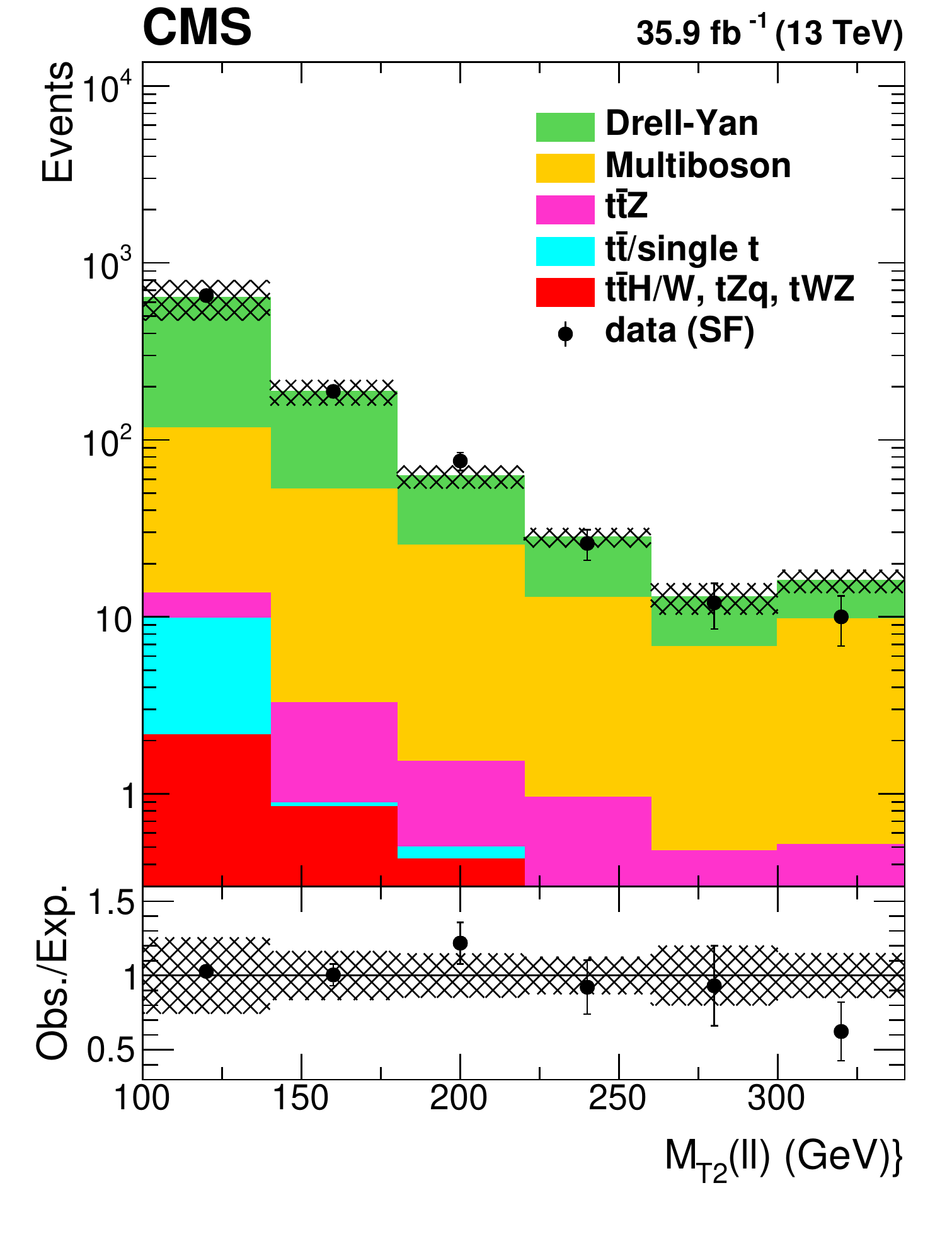}
\includegraphics[width=0.32\textwidth]{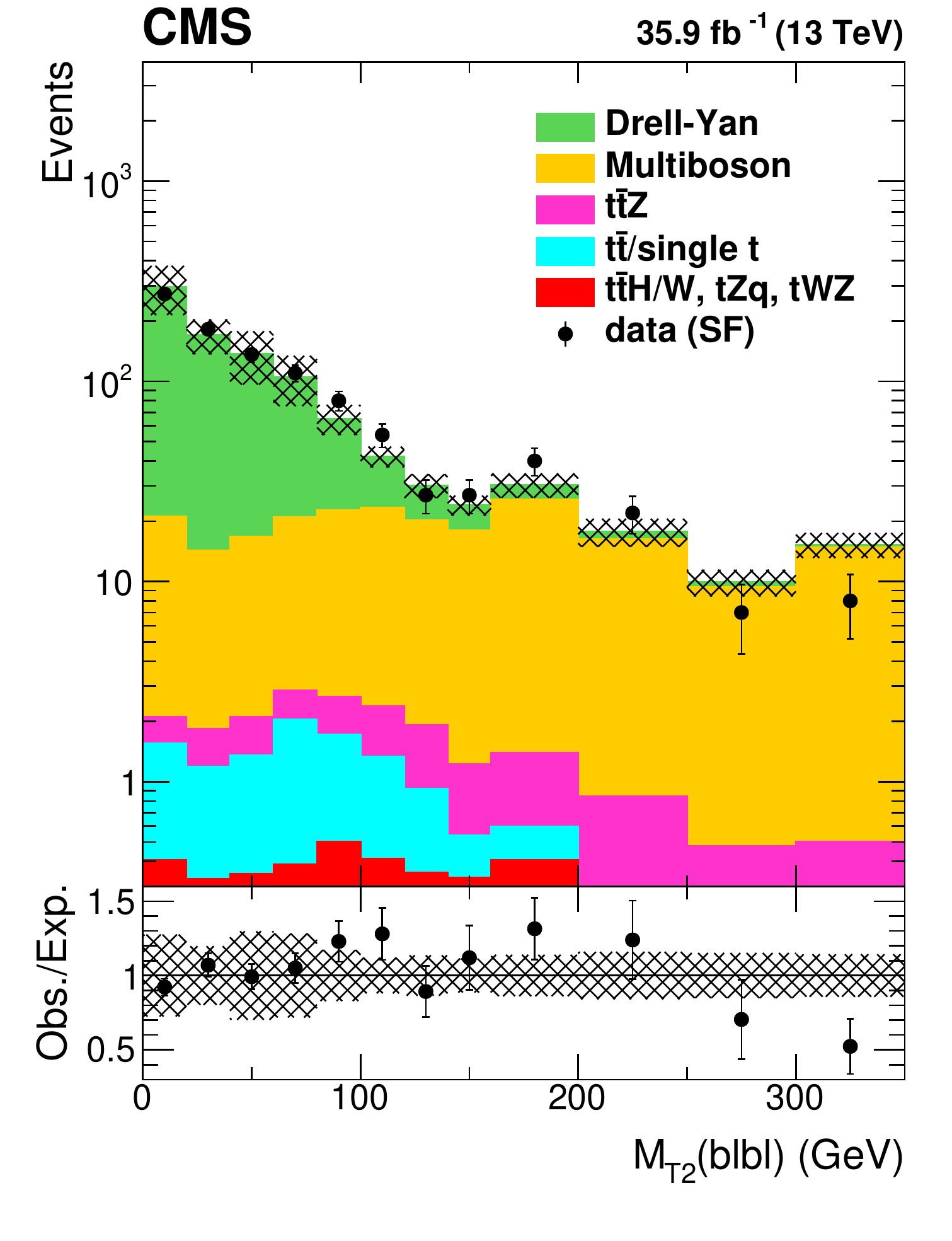}
\includegraphics[width=0.32\textwidth]{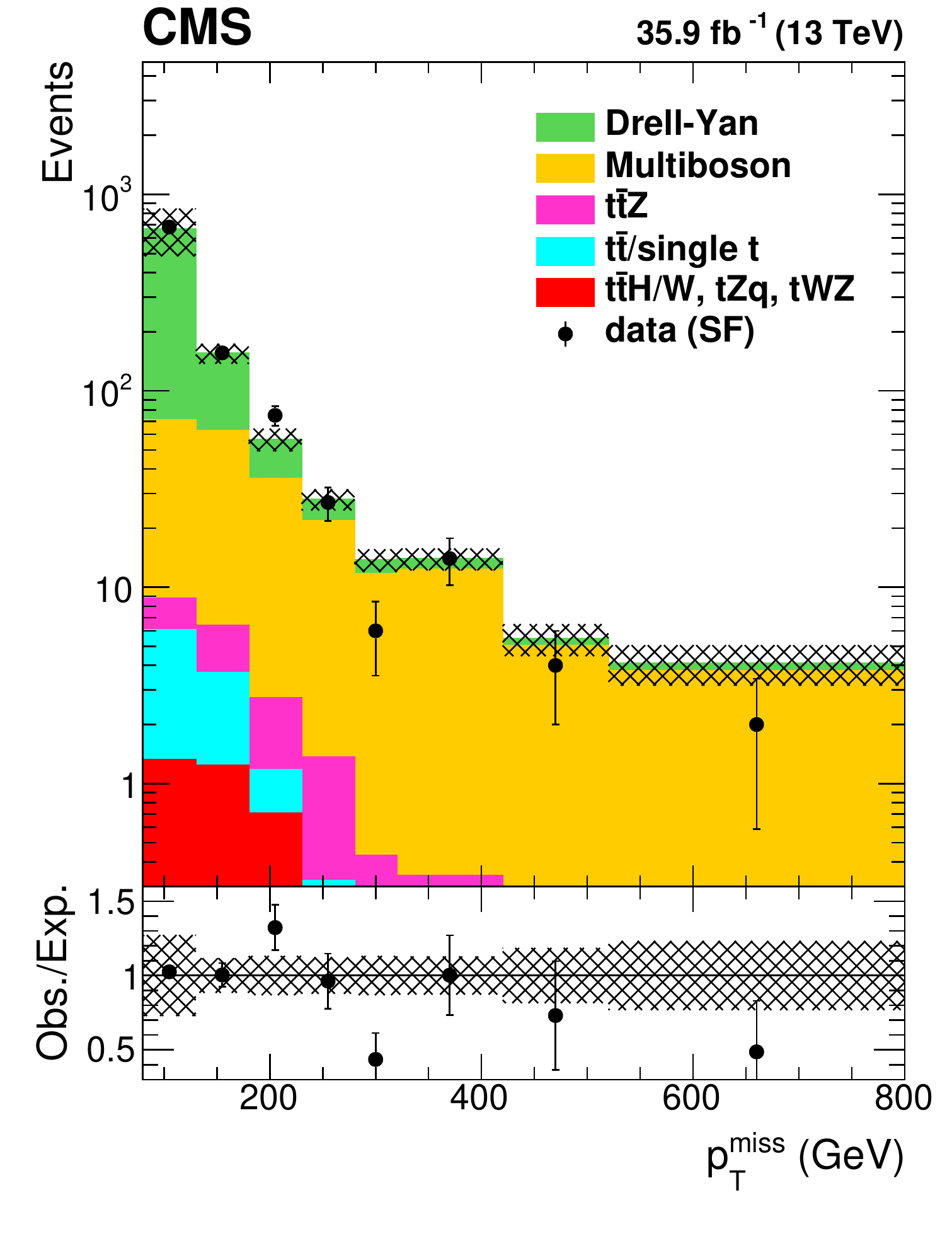}
\caption{Distributions of \mtll~(left), \mtlblb~(center), and \ptmiss~(right) for SF events falling within the \PZ boson mass window ($\abs{m(\ell\ell)-m_{\PZ}}<15\GeV$), with at least two jets and $\Nbjets=0$, $\ptmiss > 80\GeV$,
and $\mtll > 100 \GeV$.
The hatched band shows the uncertainties from experimental effects, as described in Section~\ref{sec:systematics}.}
\label{fig:DY_diboson}
\end{figure*}

\begin{figure}[htbp]
\centering
\includegraphics[width=0.49\textwidth]{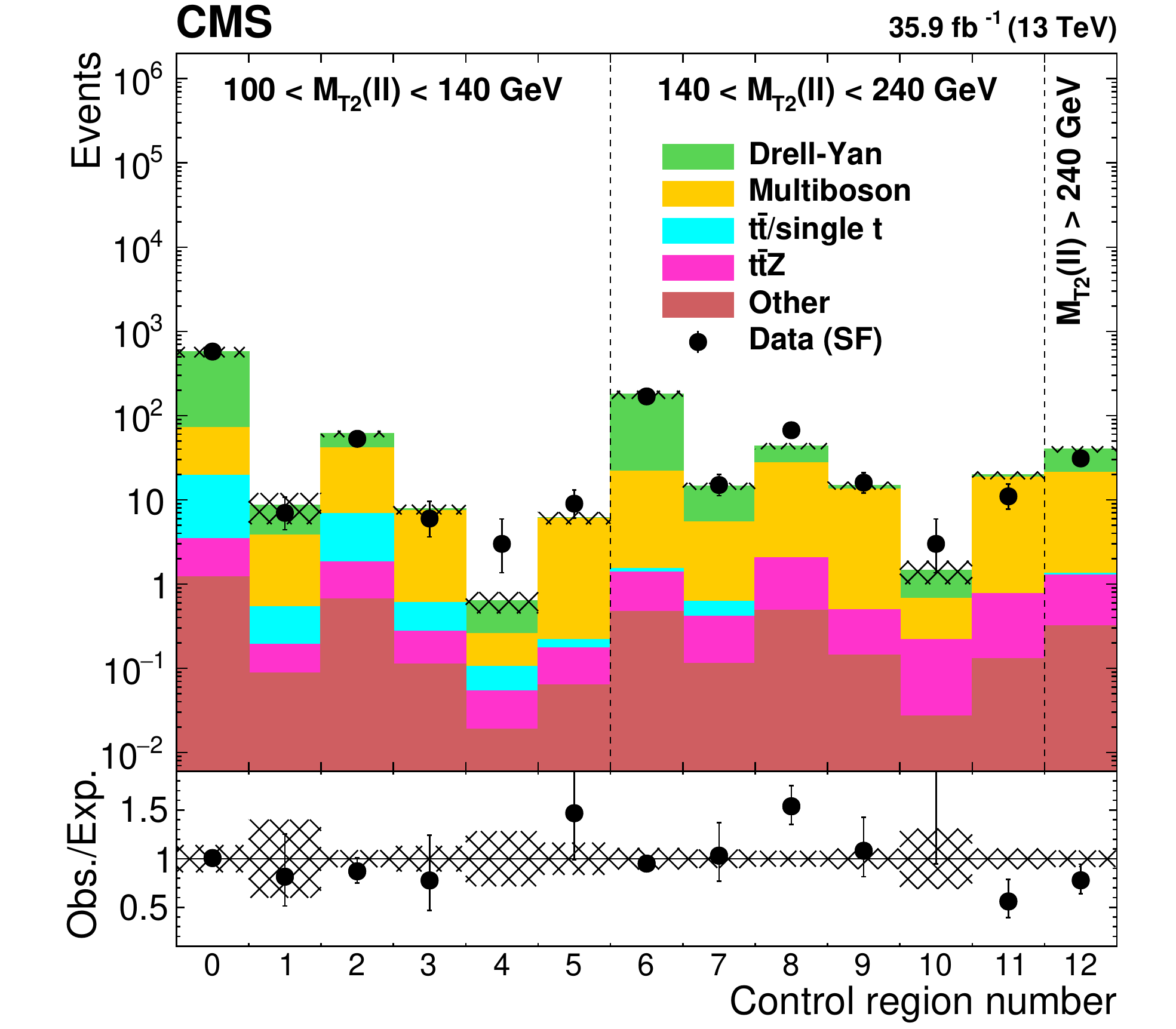}
\caption{Event yields in the 13 Drell--Yan and multiboson control regions for events with SF leptons falling within the \PZ boson mass window ($\abs{m(\ell\ell)-m_{\PZ}}<15\GeV$) and $\Nbjets=0$, after renormalizing with the scale factors obtained from the fit procedure described in the text.
The hatched band shows the uncertainties from the fit including the uncertainties from the experimental effects, as described in Section~\ref{sec:systematics}.}
\label{fig:diboson_fit}
\end{figure}

Including systematic uncertainties, we perform a likelihood fit of the predicted yields in these control regions and extract simulation-to-data scale factors that amount to $1.31 \pm 0.19$ for the Drell--Yan background and $1.19 \pm 0.17$ for the multiboson background component.
The \mtll distribution with this selection is presented in Fig.~\ref{fig:DY_diboson} (left) after applying the overall scale factors.
The fit procedure is sensitive to the Drell--Yan and multiboson contributions separately, because their \mtlblb and \ptmiss distributions differ substantially, as shown in Fig.~\ref{fig:DY_diboson} (middle) and (right), respectively.
Good agreement between the prediction and observation of both Drell--Yan and multiboson contributions is observed, and the result in all 13 control regions is shown in Fig.~\ref{fig:diboson_fit}.

\section{Systematic uncertainties and signal acceptance}
\label{sec:systematics}

Several experimental uncertainties affect the various signal and background yield estimations.
Efficiencies of the dilepton triggers, as mentioned previously, range from 95 to 99\%. The uncertainties in these efficiencies are about 1\%.
Offline lepton reconstruction and selection efficiencies are measured using $\PZ\to \ell\ell$ events in bins of lepton \pt and pseudorapidity, and as a function of the total hadronic activity in the vicinity of the lepton.
These measurements are performed separately in data and in simulation.
Typical efficiency values range from 70 to 80\%, and scale factors are used to correct the differences between data and simulation.
The uncertainties in these scale factors are less than 3\% per lepton in most of the search and control regions.

Uncertainties in the event yields resulting from the calibration of the jet energy scale are estimated by shifting the jet momenta in the
simulation up and down by one standard deviation of the jet energy corrections. Depending on the jet \pt and $\eta$,
the resulting uncertainty in the simulated yields from the jet energy scale is typically 1--5\%, except in the lowest regions in \mtll where it can be as high as 12\%.
In addition, the energy scale of deposits from soft particles that are not clustered in jets are varied within their uncertainties and the resulting uncertainty reaches 3.5\%, with
an increase up to 25\% in the lowest \mtll region.
The \PQb tagging efficiency in the simulation is corrected using scale factors determined from
data~\cite{CMS:2016kkf}, and uncertainties are propagated to all simulated events.
These contribute an uncertainty of about 1--6\% in the predicted yields depending on the transverse momentum and pseudorapidity of the b-tagged jet.

The effect of all the experimental uncertainties described above is evaluated for each of the simulated processes in all signal regions, and is considered correlated across the analysis bins and simulated processes.

Further experimental uncertainties arise from the normalization of the single top and top quark pair, Drell--Yan, and multiboson backgrounds in their respective control regions, for which uncertainties in the scale factors derived in Section~\ref{sec:backgrounds} are taken into account.
Finally, the uncertainty in the integrated luminosity is 2.5\%~\cite{lumipas}.

Several additional systematic uncertainties affect the modeling in simulation of the various processes.
Firstly, all simulated samples are reweighted according to the distribution of the true number of interactions at each bunch crossing.
The uncertainty in the total inelastic $\Pp\Pp$ cross section leads to uncertainties of 1--6\% in the expected yields.

For the \ttbar and \ttZ backgrounds, we determine the event yield changes resulting from varying the renormalization and factorization scales by a factor of two, while keeping the overall normalization from the control region in data constant.
We assign as uncertainty the envelope of the considered yield variations, treated uncorrelated between the background processes.
Uncertainties in the PDFs can have a further effect on the simulated \mtll shape.
We determine the change of acceptance in the signal regions using the PDF variations and assign the envelope of these variations---between 1 and 6\%---as a correlated uncertainty~\cite{Butterworth:2015oua}.

Measurements of the top quark \pt in \ttbar events at $\sqrt{s} = 8$ and 13\TeV show a potential mismodeling in simulation~\cite{toppt_reweighting_8TeV,Khachatryan:2016mnb}.
To evaluate the impact of this effect, we reweight the top quark \pt in the simulated \ttbar sample to match that in data, keeping the overall normalization constant.
The difference relative to the unweighted \ttbar sample is assigned as a systematic uncertainty, which typically contributes an uncertainty of about 1--2\% in the predicted yields.

For the small contribution from top quark pair production in association with a \PW{} or a Higgs boson, we take an uncertainty of 20\% in the
cross section based on the variations of the renormalization and factorization scales and the PDFs.

Finally, the statistical uncertainties due to the finite number of simulated events
are  treated as fully uncorrelated.
These maximally amount to 27\% on the rare backgrounds, with little impact on the analysis sensitivity.

A summary of the systematic uncertainties in the background prediction is presented in Table~\ref{tab:sys}.

Most of the sources of systematic uncertainties in the background estimates affect the prediction of the signal as well, and these are evaluated separately for each mass configuration of the considered simplified models.
We further estimate the effect of missing higher-order corrections for the signal acceptance by varying the renormalization and factorization scales \cite{Catani2003zt,Cacciari2003fi}
and find that uncertainties are between 1 and 19\%.
The modeling of initial-state radiation (ISR) is relevant for the SUSY signal simulation in cases where the mass difference between the top squark and the LSP is small.
The ISR reweighting is obtained in an inclusive data control region requiring an opposite-charge electron-muon pair and exactly two b jets, and is based on the number of ISR jets ($N_J^\mathrm{ISR}$) not tagged as b jets, so as to make the jet multiplicity agree with data.
The reweighting procedure is applied to SUSY Monte Carlo events and factors vary between 0.92 and 0.51 for $N_J^\mathrm{ISR}$ between 1 and 6.
We take one half of the deviation from unity as the systematic uncertainty in these reweighting factors, correlated across search regions.
It is generally found to have a small effect, but can reach 30\% for compressed mass configurations.
An uncertainty from potential differences of the modeling of \ptmiss in the fast simulation of the CMS detector with respect to data is evaluated by comparing
the reconstructed \ptmiss with the \ptmiss obtained using generator-level information. This uncertainty ranges up to 20\% and only affects the considered SUSY signal samples.
For these samples, the scale factors and uncertainties for the tagging efficiency of \PQb jets and leptons as well as the uncertainty on the modeling of pileup are evaluated separately.
For DM signal models, the uncertainty in the signal acceptance due to variations of the PDFs is considered, while for the SUSY signal models, this uncertainty was found to be redundant with the ISR uncertainty and thus not included.

\begin{table*}
  \centering
  \topcaption{Relative systematic uncertainties in the background yields in the signal regions. Where given, ranges represent the minimal and maximal changes in yield across all signal regions.}
  \label{tab:sys}
  \newcolumntype{w}{D{,}{\text{--}}{1.2}}
\begin{scotch}{lw}
  Source of systematic uncertainty & \multicolumn{1}{c}{Change in signal region yields (\%)} \\
    \hline
    Trigger efficiency & \multicolumn{1}{c}{1}  \\
    Lepton scale factors & 1,5 \\
    Jet energy scale & 1,12 \\
    Modeling of unclustered energy & 1,25 \\
    \PQb tagging & 1,6 \\
    Top quark background normalization & 3,20 \\
    \ttZ background normalization & 1,14 \\
    Multiboson background normalization & 1,8 \\
    Drell-Yan background normalization & 1,7 \\
    Integrated luminosity & \multicolumn{1}{c}{2.5} \\
    Pileup modeling & 1,6 \\
    Factorization/renormalization scales & 1,19 \\
    PDFs & 1,6 \\
    Top quark \pt modeling & 1,2 \\
    \ttX (excl. \ttZ) background normalization & 1,6 \\
    Simulated sample event count & 2,27 \\
\end{scotch}
\end{table*}

\section {Results}
\label{sec:results}
\begin{figure*}[htbp]
\centering
\includegraphics[width=0.32\textwidth]{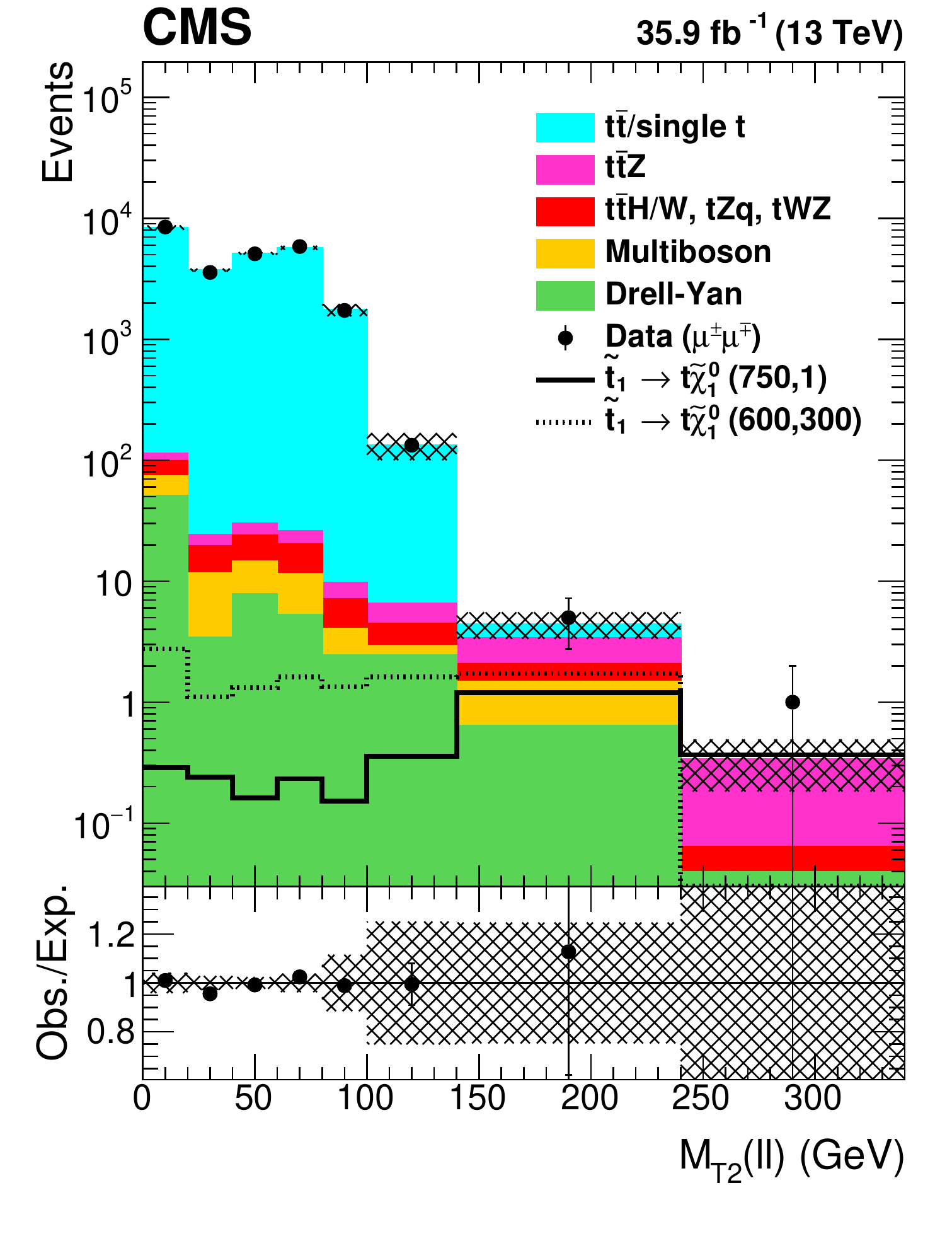}
\includegraphics[width=0.32\textwidth]{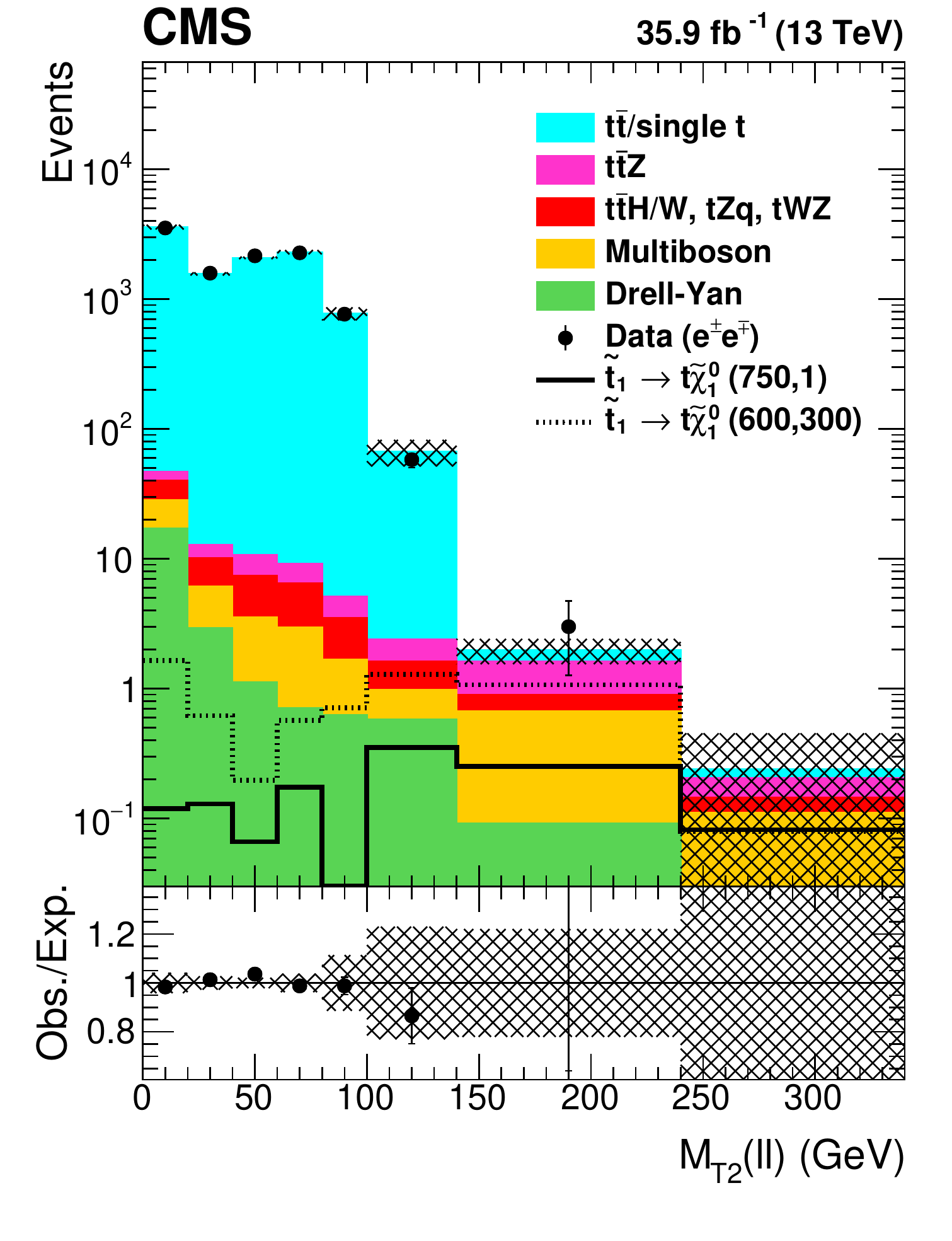}
\includegraphics[width=0.32\textwidth]{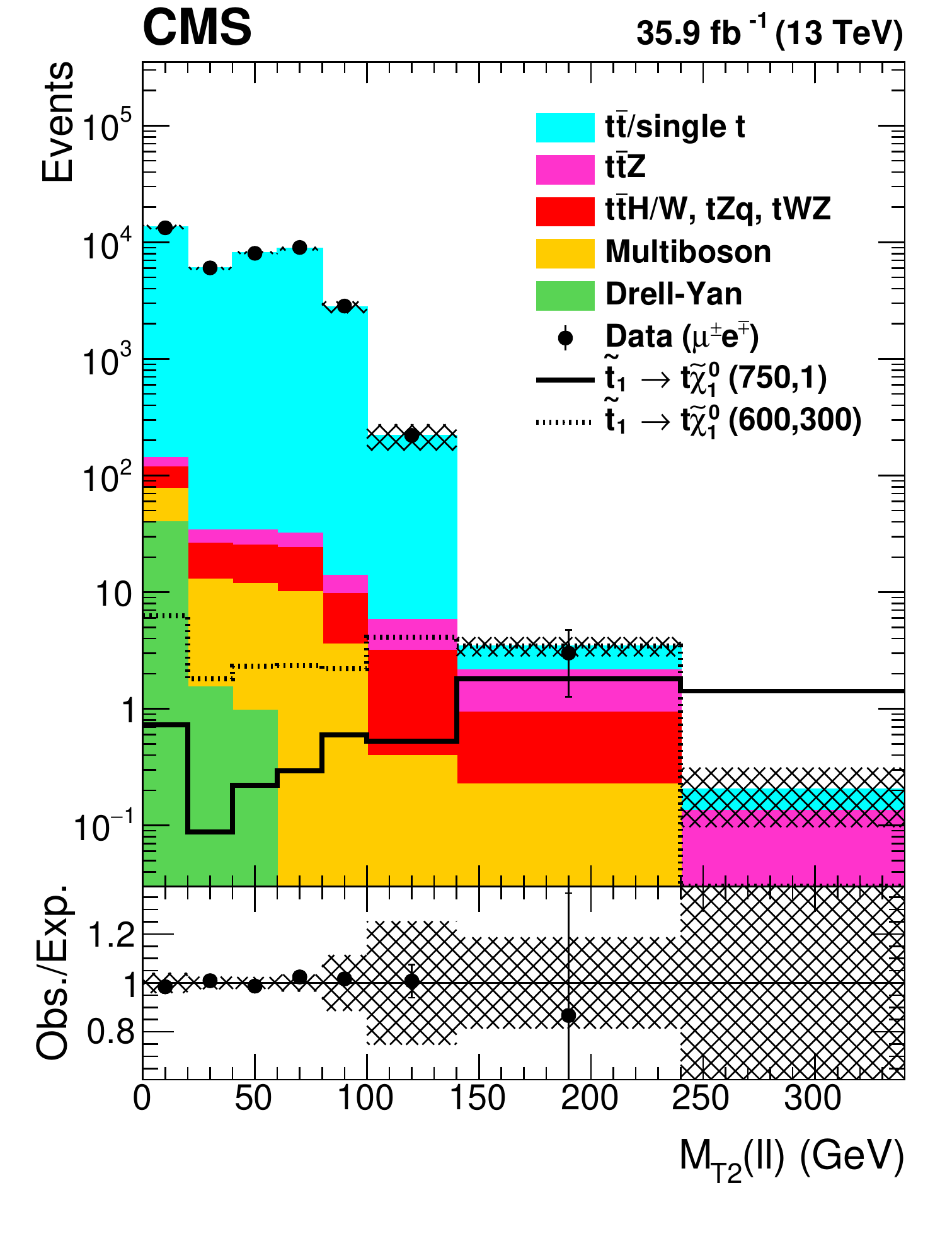}
\caption{Distributions of $\mtll$ for observed events in the $\mu\mu$ (left),  $\Pe\Pe$ (middle), and  $\Pe\mu$ (right) channels compared to the predicted SM backgrounds for the selection defined in Table~\ref{Tab:baselineSel}.
The hatched band shows the uncertainties discussed in the text.}
\label{fig:MT2ll}
\end{figure*}

\begin{figure*}[htb]
\centering
\includegraphics[width=0.32\textwidth]{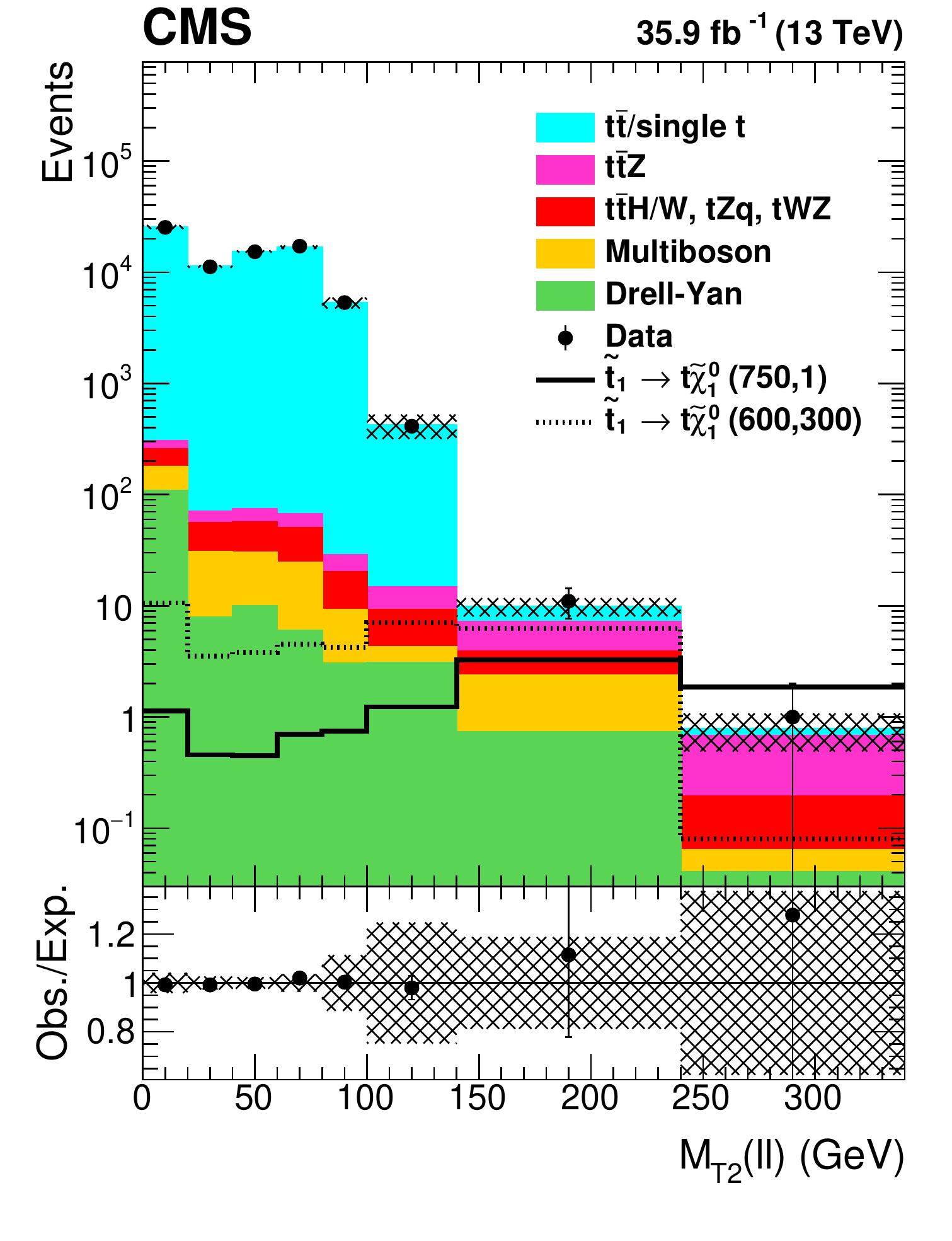}
\includegraphics[width=0.32\textwidth]{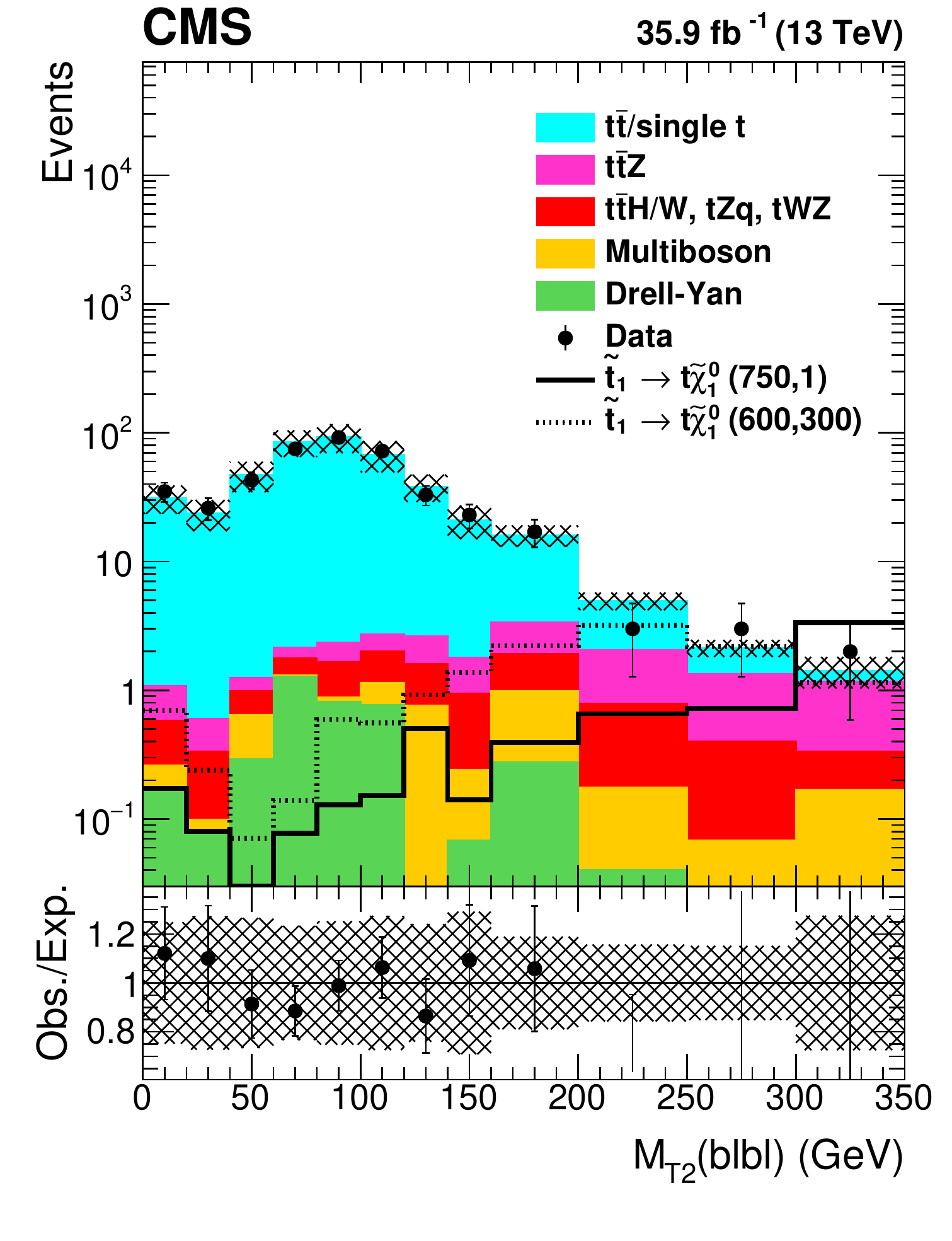}
\includegraphics[width=0.32\textwidth]{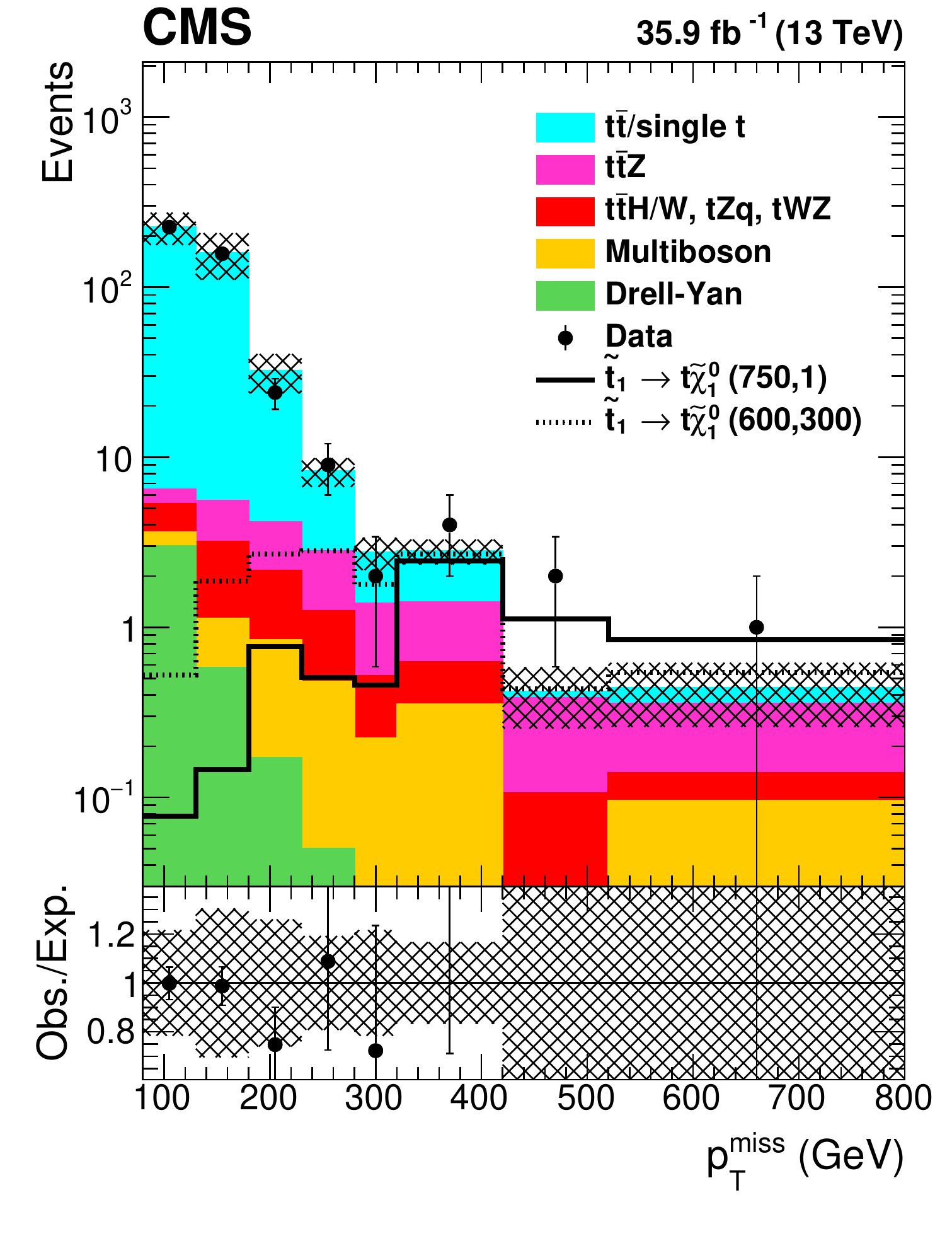}
\caption{Distributions of \mtll (left), \mtlblb (middle), and \ptmiss (right) for all lepton flavors for the selection defined in Table~\ref{Tab:baselineSel}.
Additionally, $\mtll > 100\GeV$ is required for the \mtlblb and \ptmiss distributions.
The hatched band shows the uncertainties discussed in the text.
}
\label{fig:MT2bbblbl}
\end{figure*}

\begin{figure}[htb]
\centering
\includegraphics[width=0.49\textwidth]{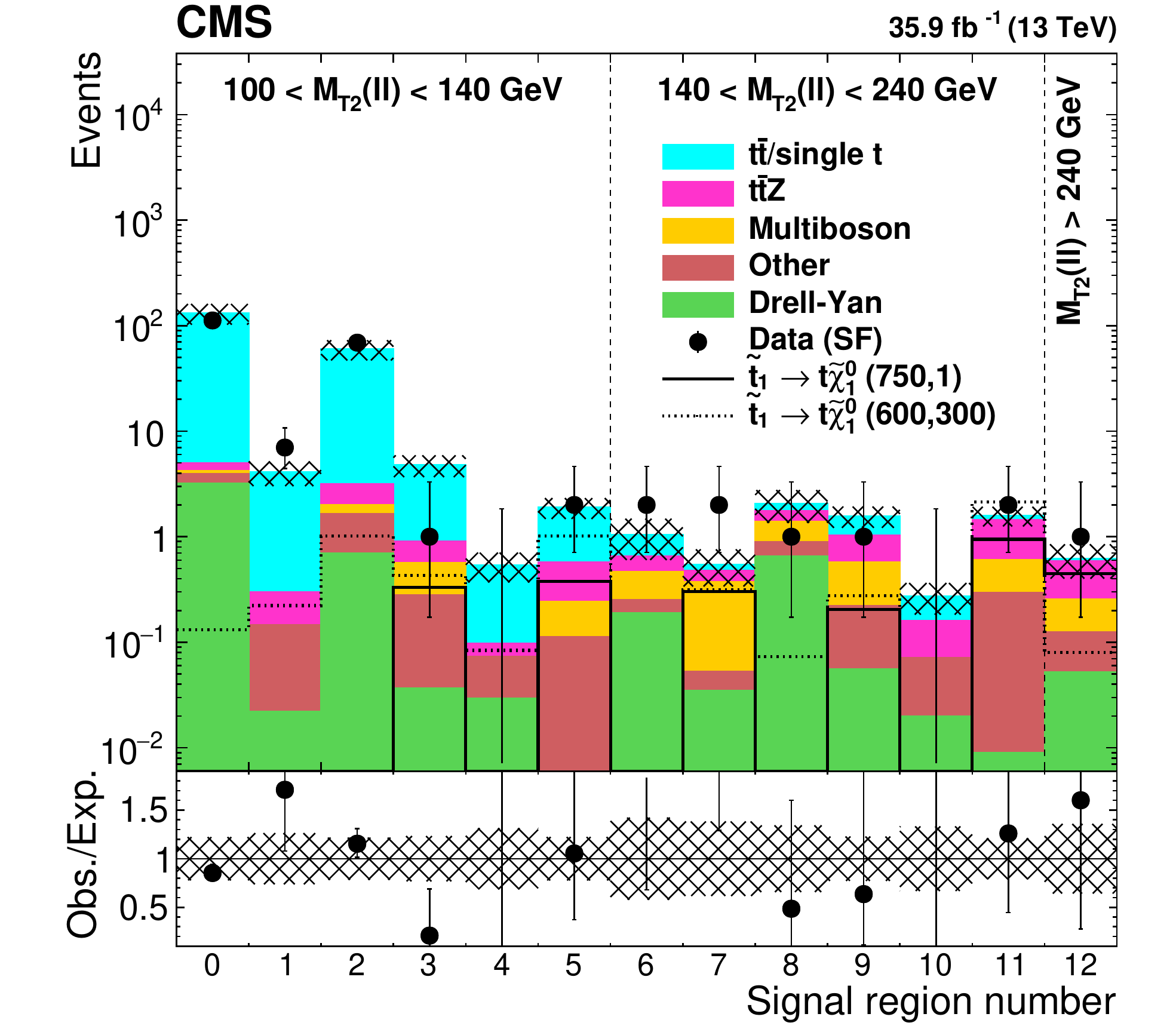}
\includegraphics[width=0.49\textwidth]{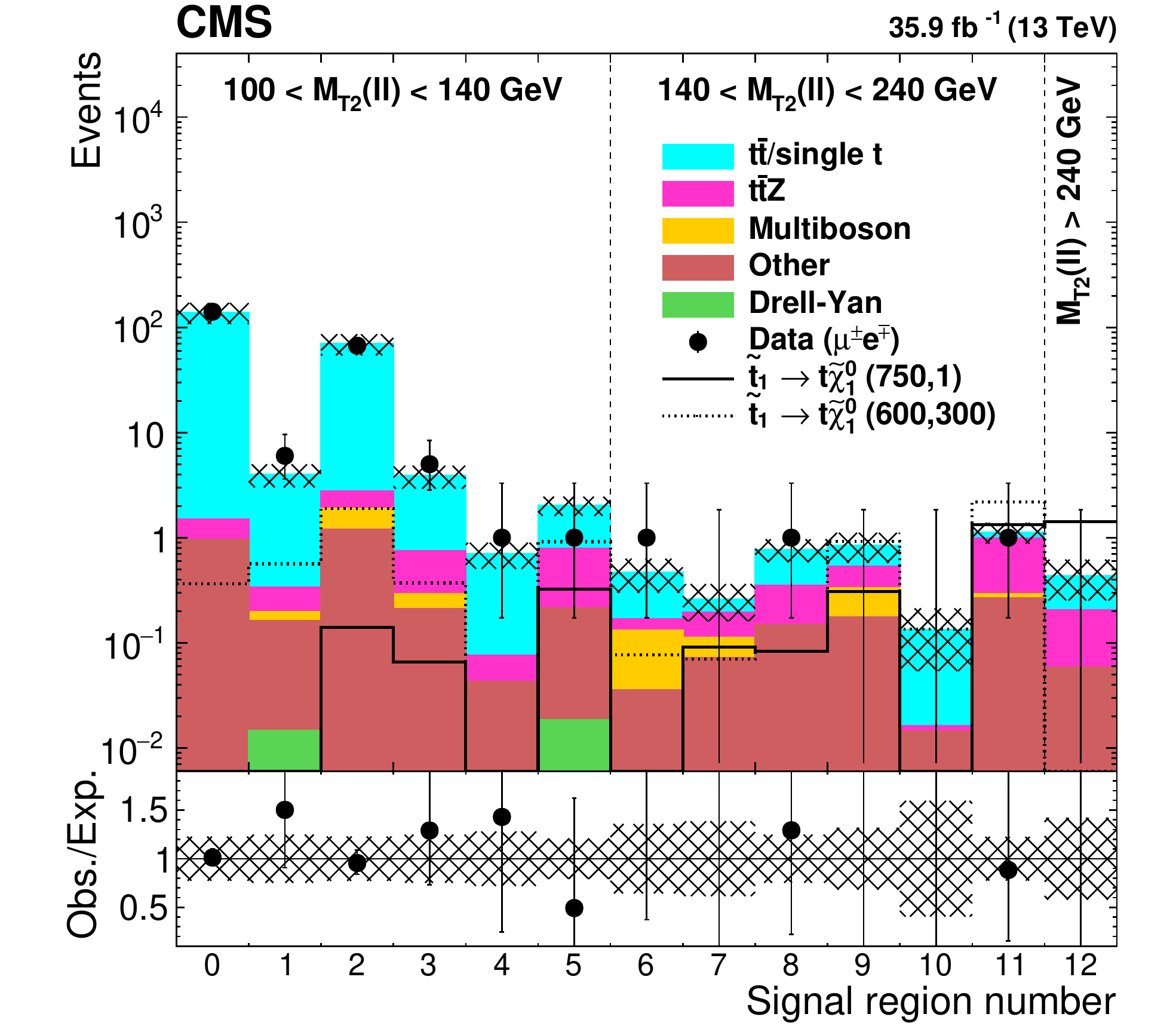}
\caption{Predicted backgrounds and observed yields in the $\Pe\Pe$ and $\mu\mu$ search regions (\cmsLeft) and the $\Pe\mu$ search regions (\cmsRight).
The hatched band shows the uncertainties discussed in the text.
}
\label{fig:SRyields}
\end{figure}

\begin{figure}[htb]
\centering
\includegraphics[width=\cmsFigWidth]{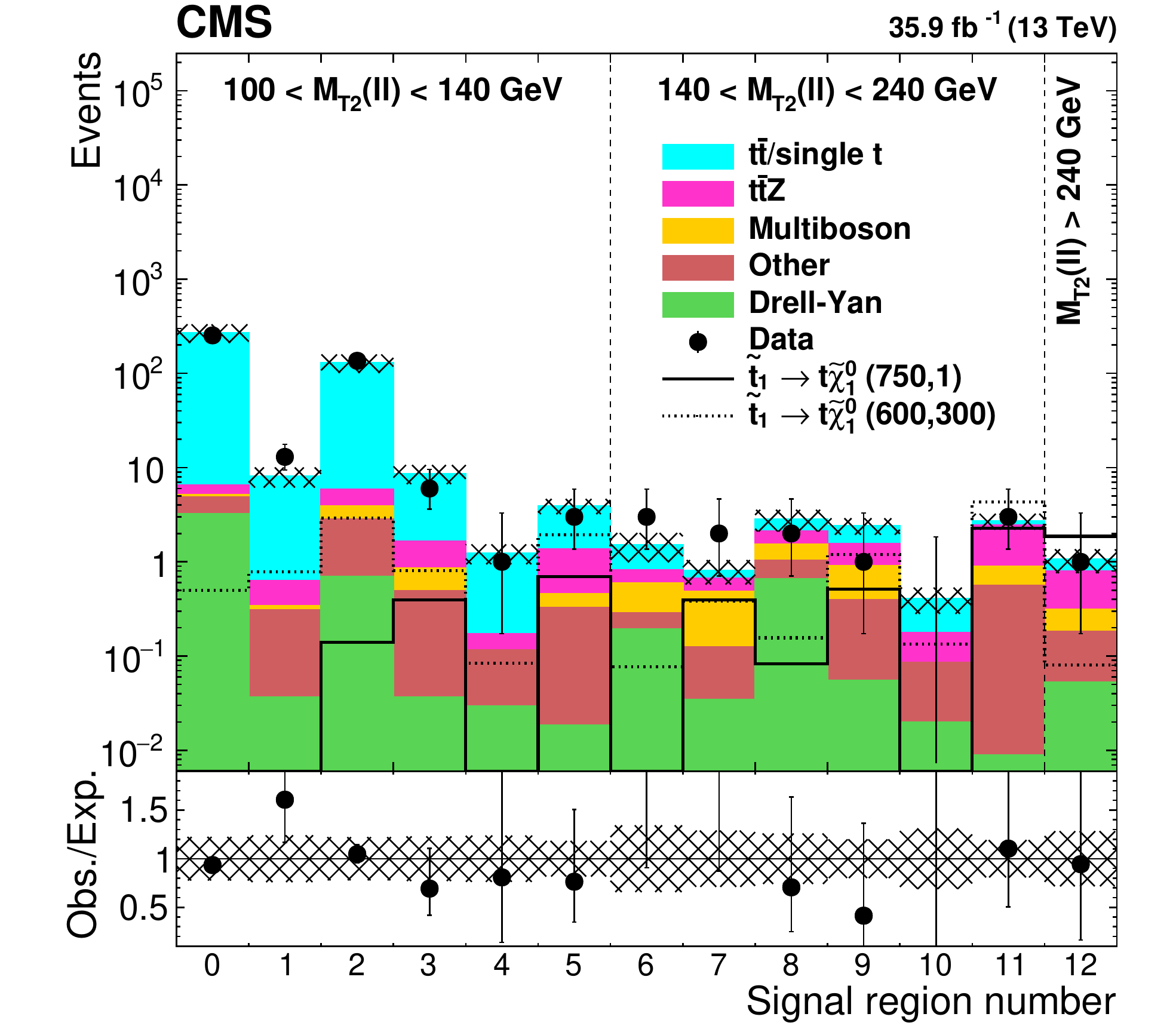}
\caption{Predicted backgrounds and observed yields in the  $\Pe\Pe$, $\mu\mu$, and $\Pe\mu$ search regions combined. The hatched band shows the uncertainties discussed in the text.}
\label{fig:SRyields_All}
\end{figure}

No significant deviation from the SM prediction is observed in any of the signal regions.
Good agreement between the predicted and observed \mtll, \mtlblb, and \ptmiss distributions is observed, as shown in Figs.~\ref{fig:MT2ll} and~\ref{fig:MT2bbblbl}, respectively.
A summary of the predicted and observed event yields for each signal region is shown in Figs.~\ref{fig:SRyields} and~\ref{fig:SRyields_All} and in Table~\ref{tab:overview}.

\begin{table*}[htb]
  \centering
  \topcaption{Total expected background and event yields in data in each of the signal regions for same-flavor ($\Pep\Pem/\PGmp\PGmm$), different-flavor ($\Pepm\PGm^\mp$), and all channels combined with all the systematic uncertainties included as described in Section~\ref{sec:systematics}.}
  \label{tab:overview}
\setlength{\tabcolsep}{2pt}
\newcolumntype{z}{D{,}{\,\pm\,}{4.3}}
\begin{scotch}{lzr{c}@{\hspace*{5pt}} zr{c}@{\hspace*{5pt}} zr}
  Signal region & \multicolumn{2}{c}{Same flavor} && \multicolumn{2}{c}{Different flavor} && \multicolumn{2}{c}{All} \\  \cline{2-3}\cline{5-6}\cline{8-9}
& \multicolumn{1}{c}{Expected} &Observed&& \multicolumn{1}{c}{Expected} & \multicolumn{1}{c}{Observed} && \multicolumn{1}{c}{Expected} &  \multicolumn{1}{c}{Observed} \\
\hline
  0  & 131,30     &         112 && 139,32     &         141 && 271,61 & 253 \\
  1  & 4.1,1.1     &           7 && 4.0,1.1    &           6 && 8.1,2.0 & 13 \\
  2  & 60 ,13      &          69 && 70 ,17     &          67 && 130,29  & 136 \\
  3  & 4.8,1.2     &           1 && 3.9,1.0    &           5 && 8.7,2.0 & 6 \\
  4  & 0.5,0.2    &           0 && 0.7,0.2    &           1 && 1.2,0.4 & 1 \\
  5  & 1.9,0.5    &           2 && 2.1,0.5    &           1 && 4.0,0.8 & 3 \\
  6  & 1.1,0.6    &           2 && 0.5,0.2    &           1 && 1.5,0.7 & 3 \\
  7  & 0.6,0.3    &           2 && 0.3,0.2    &           0 && 0.8,0.3 & 2 \\
  8  & 2.1,0.7    &           1 && 0.8,0.2    &           1 && 2.9,0.7 & 2 \\
  9  & 1.6,0.4    &           1 && 0.9,0.3    &           0 && 2.5,0.5 & 1 \\
  10 & 0.3,0.1    &           0 && 0.1,0.1    &           0 && 0.4,0.2 & 0 \\
  11 & 1.7,0.4    &           2 && 1.2,0.3    &           1 && 2.9,0.6 & 3 \\
  12 & 0.7,0.3    &          1 && 0.5,0.2    &           0 && 1.1,0.4 &  1 \\
\end{scotch}
\end{table*}

We interpret the results in the context of simplified SUSY models and combine with complementary results from the searches in the all-hadronic~\cite{stop0L2016} and the single-lepton~\cite{Sirunyan:2017xse}
final states for the \Ttt and \TbW models. Moreover, we also interpret the results in a model with DM particle pair production via a scalar or pseudoscalar mediator.

To perform the statistical interpretations, a likelihood function  is formed containing Poisson probability functions for all data regions, where the same-flavor and different-flavor signal regions are considered separately.
The control regions for the \ttZ background and for the Drell--Yan and multiboson backgrounds, as depicted in Figs.~\ref{fig:3LttZ_controlPlots} and~\ref{fig:diboson_fit}, respectively, are included as well.
The correlations of the uncertainties are taken into account as described in Section~\ref{sec:systematics}.
A profile likelihood ratio in the asymptotic approximation~\cite{Cowan:2010js} is used as the test statistic. Upper limits on the production cross section are then calculated at 95\%
confidence level (\CL) using the asymptotic $\text{CL}_{\text{s}}$ criterion~\cite{Junk1999,ClsCite}.

\begin{figure}[htbp]
\centering
\includegraphics[width=0.49\textwidth]{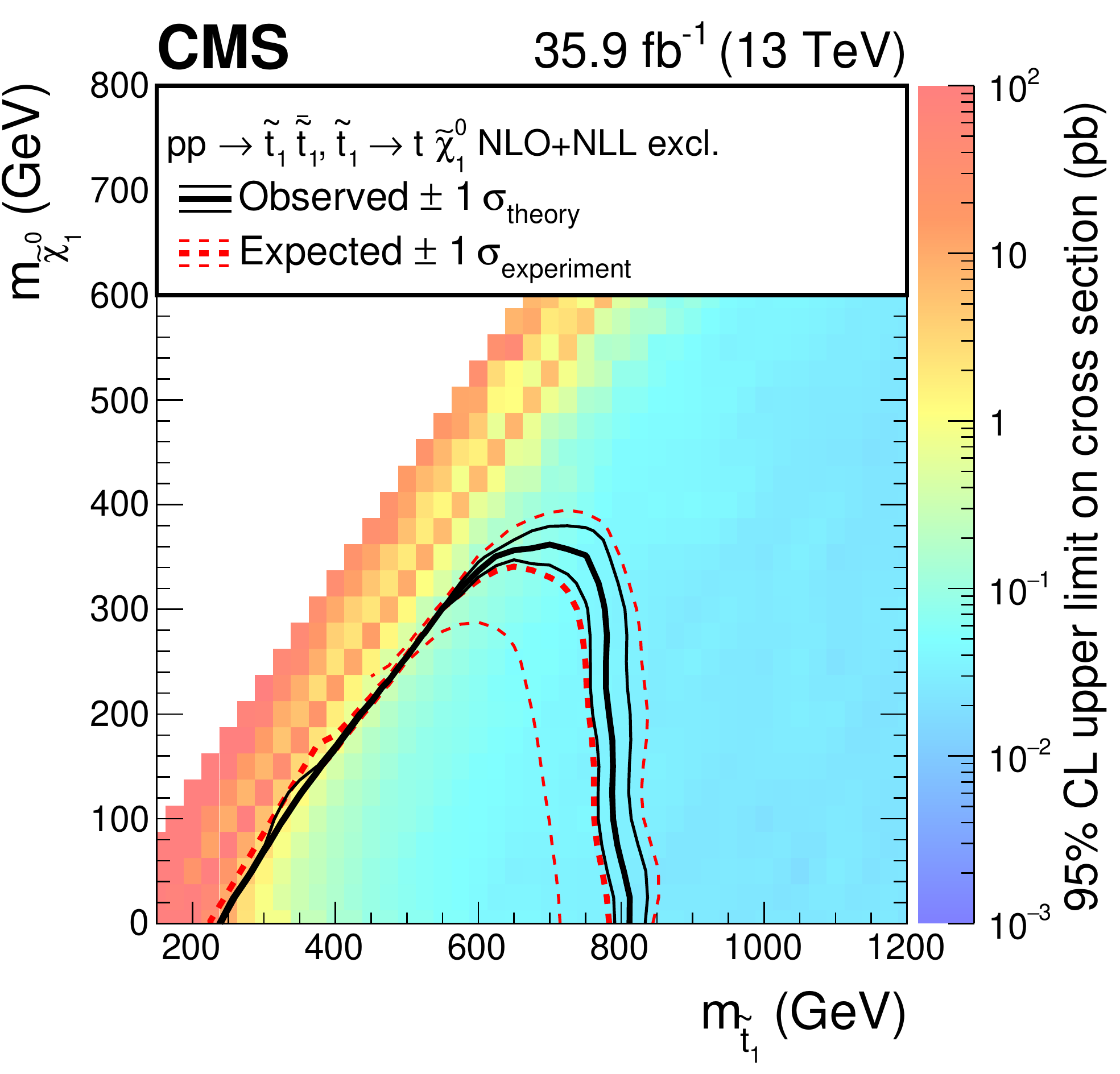}
\includegraphics[width=0.49\textwidth]{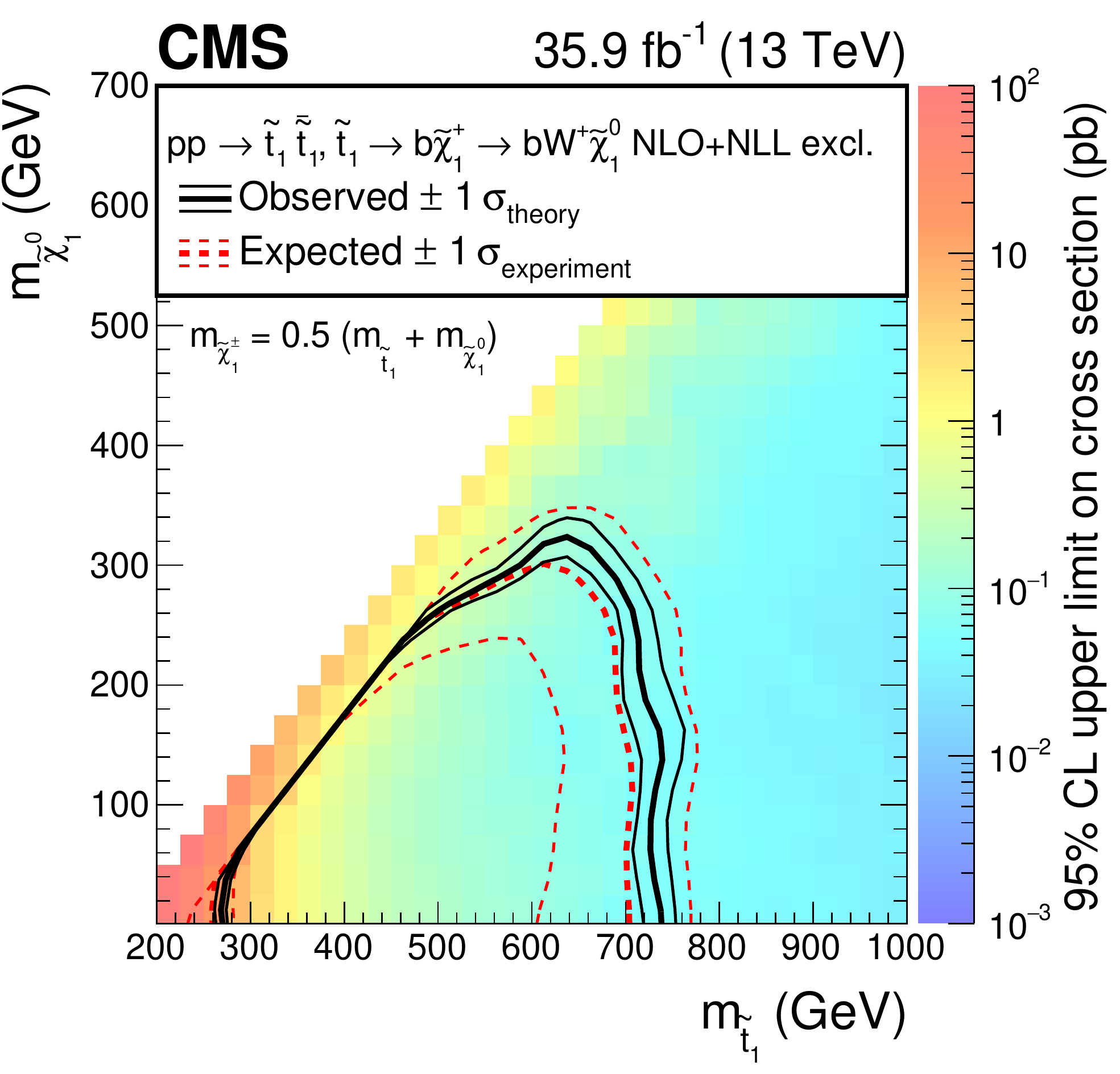}
\caption{Expected and observed limits for the \Ttt model with $\PSQtDo \to \PQt\PSGczDo$ decays (\cmsLeft) and for the \TbW model with $\PSQtDo \to \PQb\PSGcpDo \to \PQb\PWp\PSGczDo$ decays (\cmsRight) in the $m_{\PSQtDo}$--$m_{\PSGczDo}$ mass plane.
The color indicates the 95\% \CL upper limit on the cross section times the square of the branching fraction at each point in the plane.
The area below the thick black curve represents
the observed exclusion region at 95\% \CL assuming 100\% branching fraction,
while the dashed red lines indicate the expected limits at 95\% \CL and the region containing 68\% of the distribution of limits expected under the background-only hypothesis.
The thin black lines show the effect of the theoretical uncertainties in the signal cross section.}
\label{fig:limit_T2tt}
\end{figure}

\begin{figure*}[htb]
\centering
\includegraphics[width=0.45\textwidth]{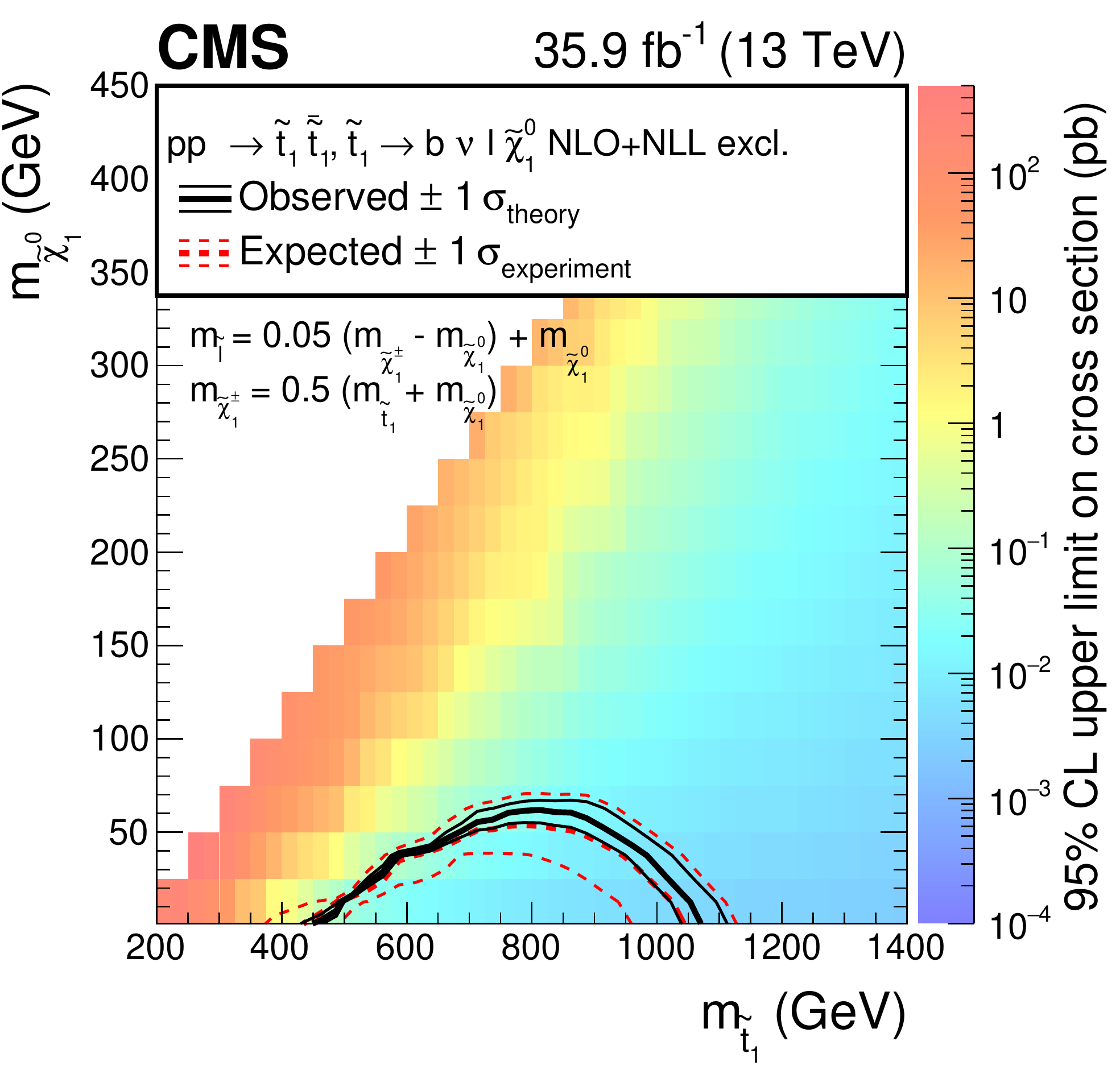}
\includegraphics[width=0.45\textwidth]{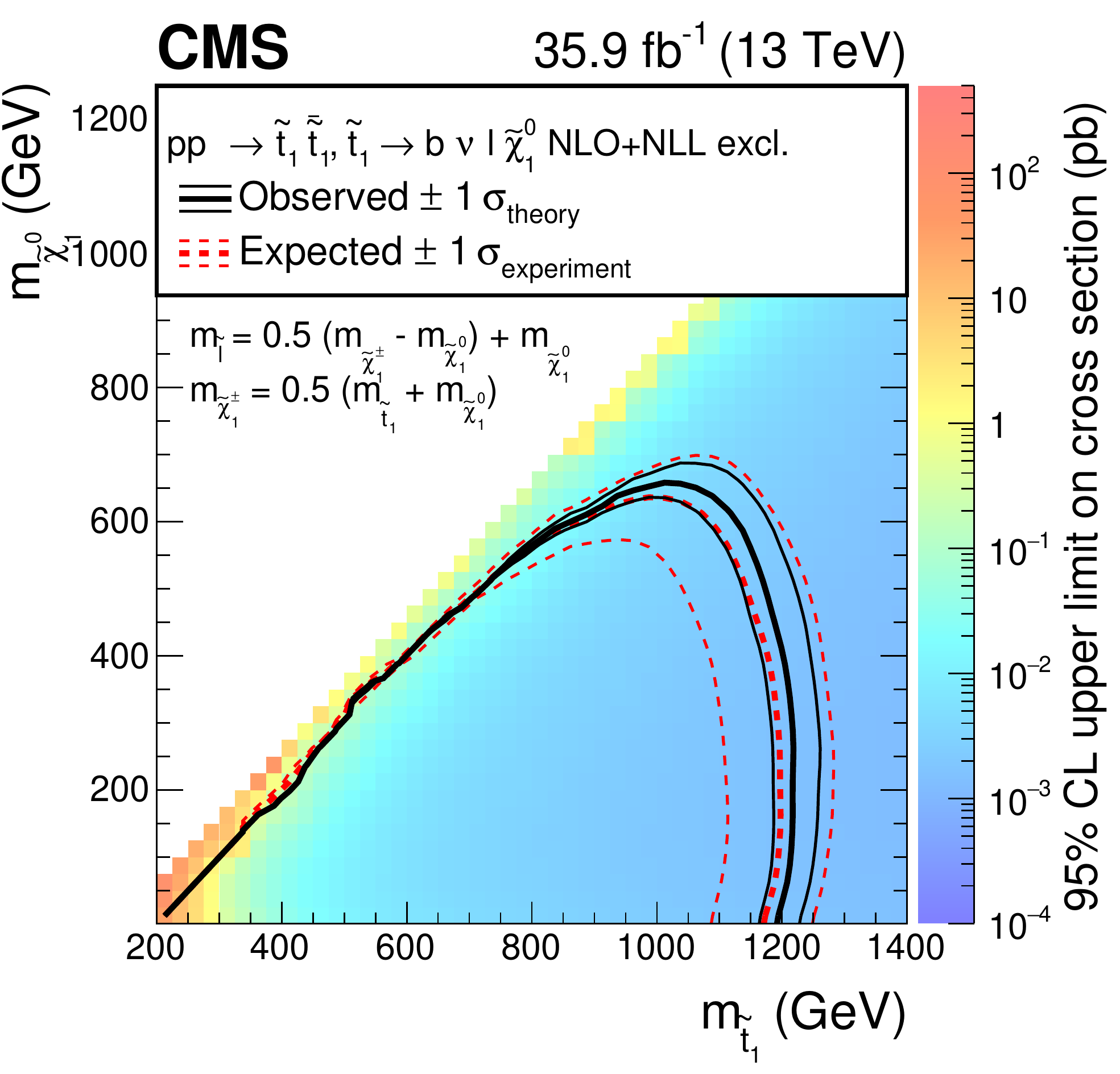}
\includegraphics[width=0.45\textwidth]{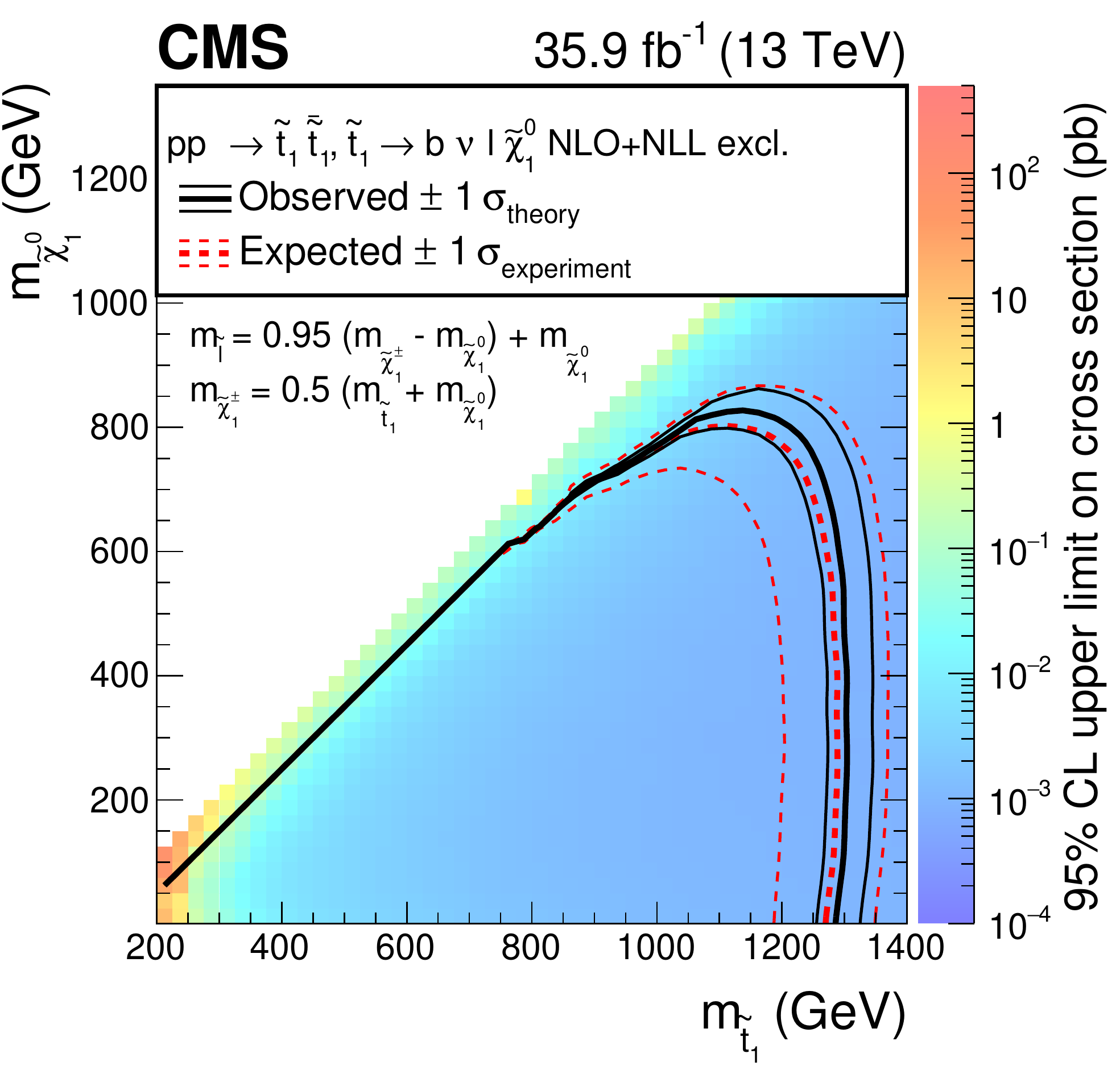}
\caption{Expected and observed limits for the \Tbbllnunu model with $\PSQtDo \to \PQb\PSGcpDo \to \PQb\PGn\tilde{\ell} \to \PQb\PGn\ell\PSGczDo$ decays in the $m_{\PSQtDo}$--$m_{\PSGczDo}$ mass plane for
three different mass configurations defined by $m_{\tilde{\ell}} = x \, (m_{\PSGcpDo} - m_{\PSGczDo}) + m_{\PSGczDo}$ with $x=0.05$ (upper left), $x=0.5$ (upper right), and $x=0.95$ (lower).
The description of curves is the same as in the caption of Fig.~\ref{fig:limit_T2tt}.}
\label{fig:limit_T8bbllnunu}
\end{figure*}

The SUSY interpretations are given in the $m_{\PSQtDo}$--$m_{\PSGczDo}$ plane in Figs.~\ref{fig:limit_T2tt} and~\ref{fig:limit_T8bbllnunu}.
The color on the z axis indicates the 95\% \CL upper limit
on the cross section times the square of the branching fraction at each point in the $m_{\PSQtDo}$--$m_{\PSGczDo}$ plane.
The area below the thick black curve represents the observed exclusion region at 95\% \CL
assuming 100\% branching fraction,
while the dashed red lines indicate the expected limit at 95\% \CL and the region containing 68\% of the distribution of limits expected under the background-only hypothesis.
The thin black lines show the effect of the theoretical uncertainties in the signal cross section.
In the \Ttt model we exclude mass configurations with $m_{\PSGczDo}$ up to 360\GeV and  $m_{\PSQtDo}$ up to 800\GeV, assuming that the top quarks are unpolarized.
Because this choice may have a significant impact on the kinematic properties of the final state particles~\cite{Belanger:2012tm}, we also check that for purely right-handed polarization, the limit increases by about 50\GeV in both $m_{\PSQtDo}$ and $m_{\PSGczDo}$, while for purely left-handed polarization, the limit
decreases by about 50\GeV in $m_{\PSQtDo}$ and by 70\GeV in $m_{\PSGczDo}$.

The results for the \TbW and \Tbbllnunu models are shown in Figs.~\ref{fig:limit_T2tt} (right) and~\ref{fig:limit_T8bbllnunu}.
We exclude mass configurations with $m_{\PSGczDo}$ up to 320\GeV and  $m_{\PSQtDo}$ up to 750\GeV in the \TbW model.
The sensitivity in the \Tbbllnunu model strongly depends on the intermediate slepton mass and is largest when $x = 0.95$ in $m_{\tilde{\ell}} = x \, (m_{\PSGcpDo} - m_{\PSGczDo}) + m_{\PSGczDo}$.
In this case, excluded masses reach up to 800\GeV for $m_{\PSGczDo}$ and 1300\GeV for $m_{\PSQtDo}$. These numbers reduce to 660\GeV for $m_{\PSGczDo}$ and 1200\GeV for $m_{\PSQtDo}$ when $x=0.5$
and to 50\GeV for $m_{\PSGczDo}$ and 1000\GeV for $m_{\PSQtDo}$ when $x=0.05$.

\begin{figure}[htb]
\centering
\includegraphics[width=\cmsFigWidth]{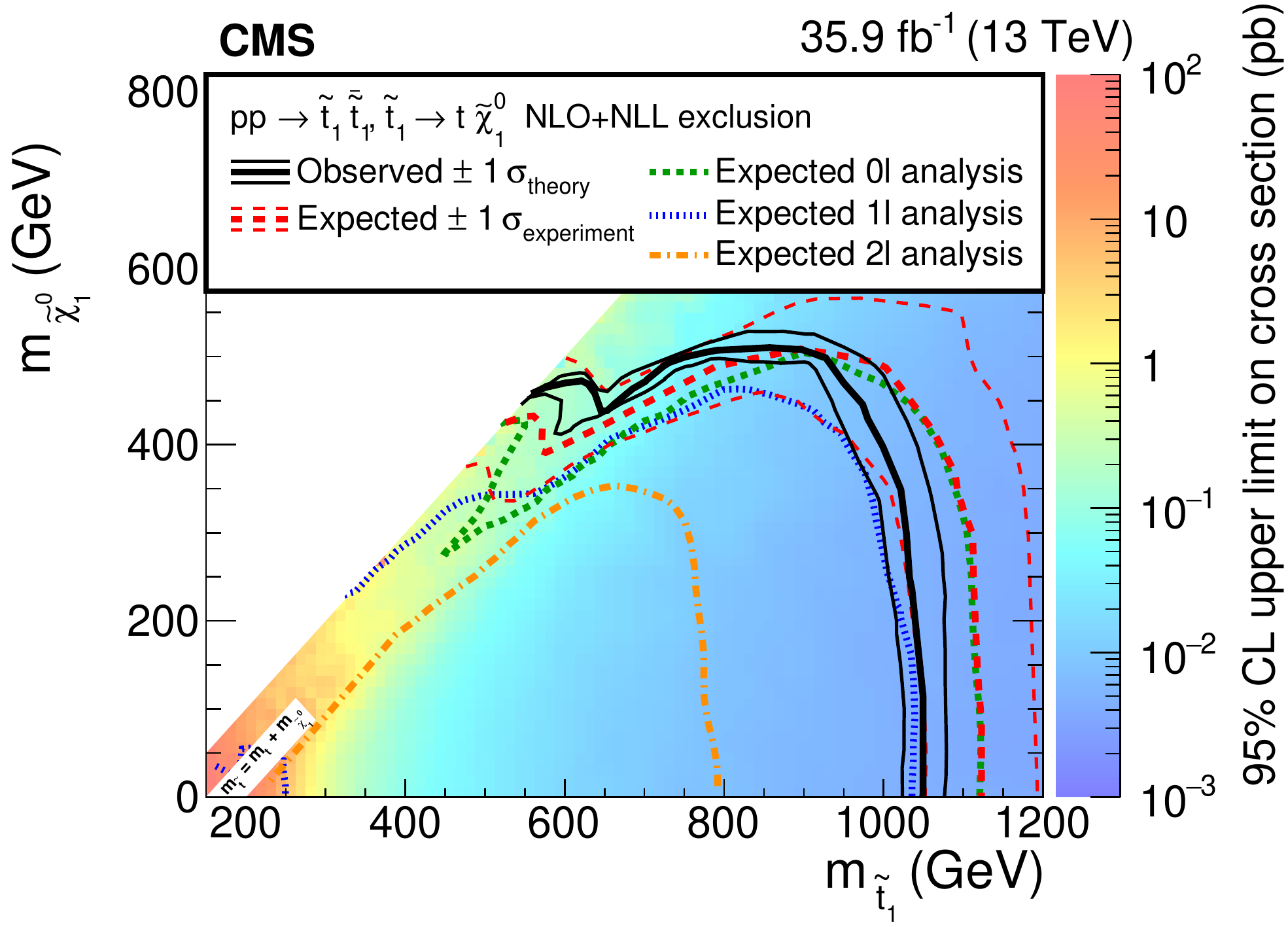}
\caption{Expected and observed limits for the \Ttt model with $\PSQtDo \to \PQt\PSGczDo$ decays in the $m_{\PSQtDo}$--$m_{\PSGczDo}$ mass plane combining the dilepton final state with the single-lepton~\cite{Sirunyan:2017xse} and the all-hadronic~\cite{stop0L2016} final states as described in the text.
The color indicates the 95\% \CL upper limit on the cross section times the square of the branching fraction at each point in the plane.
The area below the thick black curve represents
the observed exclusion region at 95\% \CL assuming 100\% branching fraction,
while the dashed red lines indicate the expected limits at 95\% \CL and the region containing 68\% of the distribution of limits expected under the background-only hypothesis.
The thin black lines show the effect of the theoretical uncertainties in the signal cross section.
The green short-dashed, blue dotted, and long-short-dashed orange curves show the expected individual limits for the all-hadronic, single-lepton, and dilepton analyses, respectively.
The whited out area on the diagonal corresponds to configurations where the mass difference between \PSQtDo and \PSGczDo is very close to the top quark mass.
In this region the signal acceptance strongly depends on the \PSGczDo mass and is therefore hard to model.
}
\label{fig:limits:T2ttComb}
\end{figure}

\begin{figure}[htb]
\centering
\includegraphics[width=\cmsFigWidth]{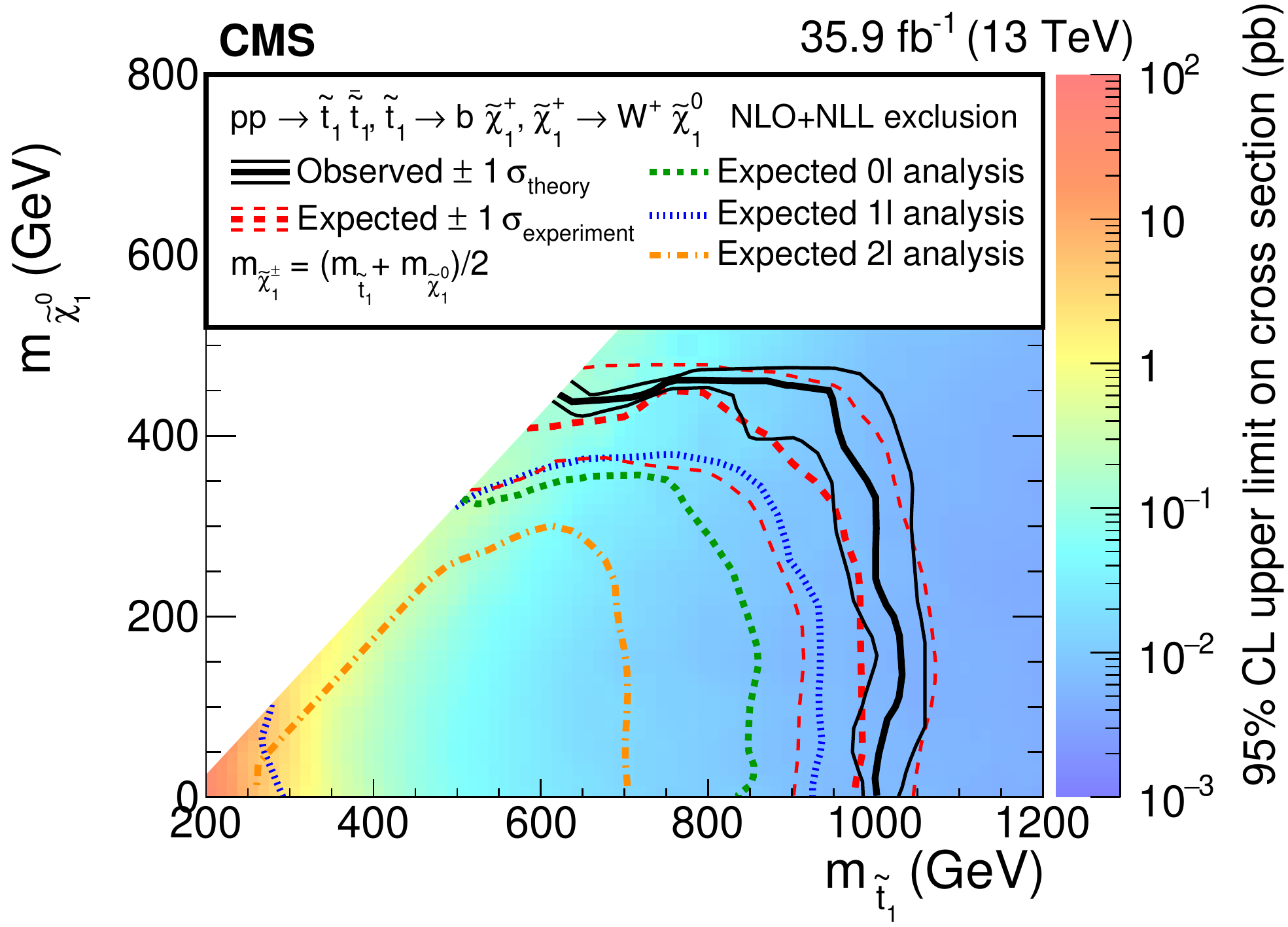}
\caption{Expected and observed limits for the \TbW model with $\PSQtDo \to \PQb\PSGcpDo \to \PQb\PWp\PSGczDo$ decays in the $m_{\PSQtDo}$--$m_{\PSGczDo}$ mass plane combining the dilepton final state with the all-hadronic~\cite{stop0L2016} and the single-lepton~\cite{Sirunyan:2017xse} final states as described in the text.
The mass of the chargino is chosen to be $(m_{\PSQtDo} + m_{\PSGczDo})/2$.
The description of curves is the same as in the caption of Fig.~\ref{fig:limits:T2ttComb}.}
\label{fig:limits:T2bwComb}
\end{figure}

Besides the dilepton search described in this paper, searches for direct top squark pair production were also performed in final states with a single lepton~\cite{Sirunyan:2017xse} and without leptons~\cite{stop0L2016}.
The signal and control regions for these two searches and the dilepton search are mutually exclusive.
A statistical combination of the results of the three searches is performed in the context of the \Ttt and \TbW scenarios of top squark pair production,
taking into account correlations in both signal and expected
background yields in the different analyses.
Figure~\ref{fig:limits:T2ttComb} shows the combination of the results of the three searches
for direct top squark pair production for the \Ttt model with $\PSQtDo\to\PQt\PSGczDo$ decays.
The combined result excludes a top squark mass of 1050\GeV for a massless LSP, and an LSP mass of 500\GeV for a top squark mass of 900\GeV.
The combination is driven primarily by the all-jet search, except in the region of small mass splitting between the top squark and the LSP
where searches in the zero- and one-lepton channels have similar sensitivity.
Figure~\ref{fig:limits:T2bwComb} shows the equivalent limits for direct top squark pair production for the \TbW model with $\PSQtDo\to\PQb\PSGcpDo$, $\PSGcpDo\to\PWp\PSGczDo$ decays.
The combined result for this scenario excludes a top squark mass of 1000\GeV for a massless LSP and an LSP mass of 450\GeV for a top squark mass of 900\GeV.
The combination extends the sensitivity to both top squark and LSP masses by about 50\GeV compared to the most sensitive individual result coming from the one-lepton channel.

\begin{figure}[htbp]
  \centering
  \includegraphics[width=0.45\textwidth]{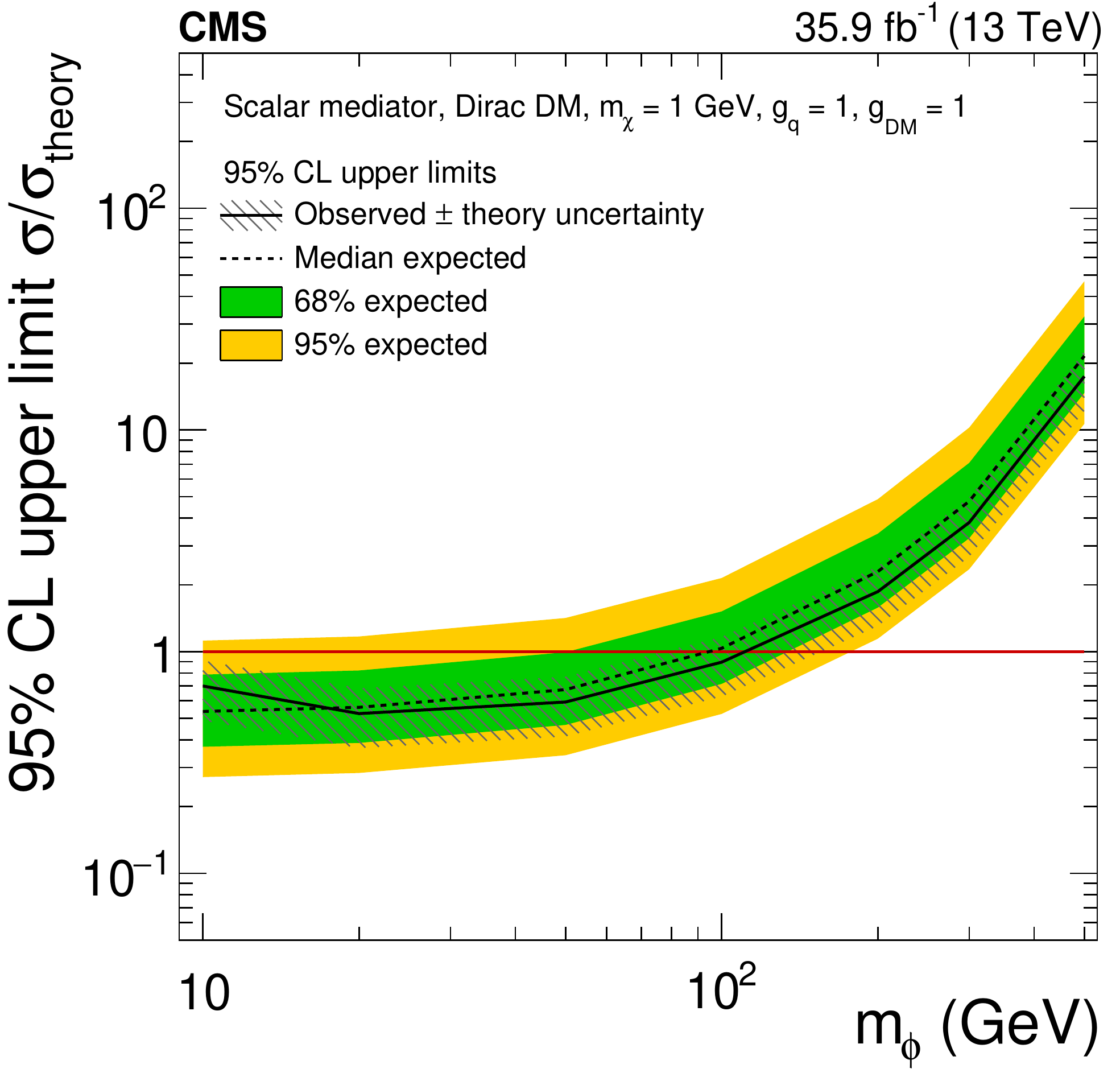}
  \includegraphics[width=0.45\textwidth]{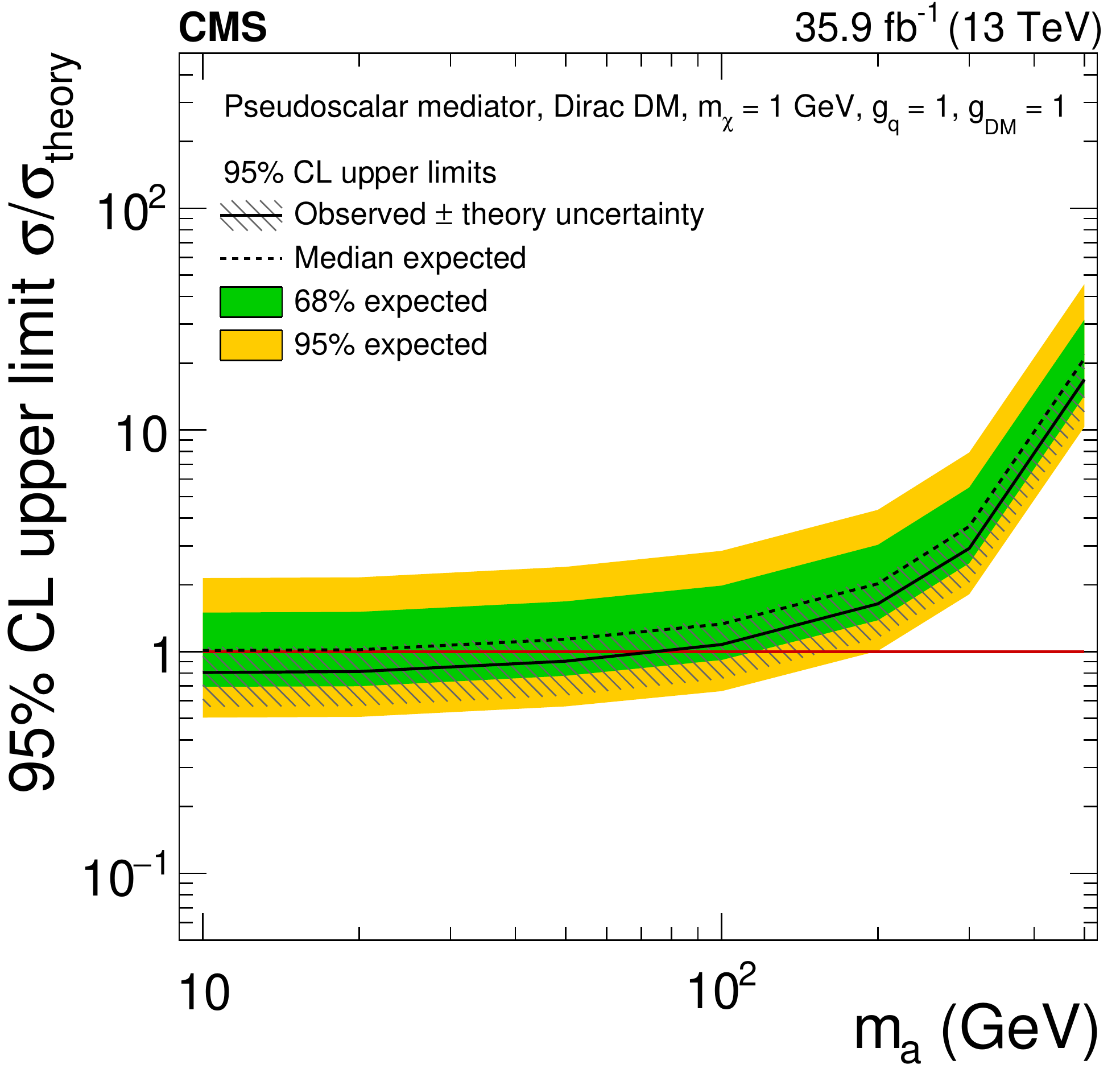}
  \caption{The 95\% \CL expected (dashed line) and observed limits (solid line) on $\mu=\sigma/\sigma_{\text{theory}}$ for a fermionic DM particle with $m_{\PGc}=1\GeV$ assuming different scalar (\cmsLeft) and pseudoscalar (\cmsRight) mediator masses.
           The green and yellow bands represent the regions containing 68 and 95\%, respectively, of the distribution of limits expected under the background-only hypothesis.
           The horizontal red line indicates $\mu= 1$.
           The mediator couplings are set to $g_{\PQq}=g_\mathrm{DM}=1$.
           The gray hashed band around the observed limit corresponds to a 30\% theory uncertainty in the inclusive signal cross section.
           }
  \label{fig:DMresults}
\end{figure}

Limits on the production of DM particle pairs in association with top quark pairs via a scalar or pseudoscalar mediator are listed in Table~\ref{tab:DMresults}, assuming $g_{\PQq}=g_\mathrm{DM}=1$.
The results are presented as ratios $\mu=\sigma/\sigma_{\text{theory}}$ of the 95\% \CL expected and observed upper limits on the cross section $\sigma$ with respect to the simplified model cross section expectations $\sigma_{\rm theory}$.
Results are shown for different DM particle and mediator masses, and for both scalar and pseudoscalar mediators.
Figure~\ref{fig:DMresults} shows expected and observed limits as a function of the mediator mass for DM particles $\PGc$ with a mass of 1\GeV.
We exclude scalar mediators with masses up to 100\GeV and pseudoscalar mediators with masses up to 50\GeV.

\begin{table*}[tb]
\centering
\renewcommand*{\arraystretch}{1.3}
 \topcaption{
   Ratios $\mu=\sigma/\sigma_{\text{theory}}$ of the 95\% \CL expected and observed limits to simplified model expectations for different DM particle masses and mediator masses and for scalar ($\phi$) and pseudoscalar ($\mathrm{a}$) mediators under the assumption $g_{\PQq}=g_\mathrm{DM}=1$.
The uncertainties reflect the 68\% band of the expected limits.
}
  \label{tab:DMresults}
 \begin{scotch}{rrc{c}ccc{c}@{\hspace*{5pt}} cc}
$m_{\PGc}$& $m_{\phi/\Pa}$&\multicolumn{3}{c}{Scalar}&&\multicolumn{3}{c}{Pseudoscalar}\\ \cline{3-5}\cline{7-9}
 (\GeVns{}) & (\GeVns{}) & $\sigma_{\text{theory}}$ (fb) & Expected & Observed && $\sigma_{\text{theory}}$ (fb) & Expected & Observed \\
\hline
1         & 10    & 21357 & $0.54^{+0.25}_{-0.16}$ & 0.70  && 451 & $1.01^{+0.49}_{-0.32}$ & 0.81    \\
1         & 20    & 10954 & $0.56^{+0.26}_{-0.17}$ & 0.53  && 411 & $1.02^{+0.49}_{-0.32}$ & 0.81    \\
1         & 50    & 3086  & $0.67^{+0.32}_{-0.21}$ & 0.59  && 308 & $1.14^{+0.55}_{-0.36}$ & 0.91    \\
1         & 100   & 720   & $1.04^{+0.48}_{-0.32}$ & 0.90  && 193 & $1.33^{+0.65}_{-0.42}$ & 1.08    \\
1         & 200   & 101   & $2.30^{+1.11}_{-0.72}$ & 1.87  && 87.8& $2.02^{+1.01}_{-0.64}$ & 1.64    \\
1         & 300   & 30.5  & $4.8^{+2.3}_{-1.5}$    & 3.8   && 39.5& $3.7^{+1.8}_{-1.2}$    & 2.9     \\
1         & 500   & 4.95  & $21.6^{+10.9}_{-6.9}$  & 17.4  && 5.14& $21.0^{+10.4}_{-6.7}$  & 16.9    \\[\cmsTabSkip]
10        & 10    & 101   & $18.8^{+8.8}_{-5.8}$   & 16.6  && 15.2& $19.3^{+9.3}_{-6.1}$   & 15.3    \\
10        & 15    & 127   & $17.0^{+8.0}_{-5.2}$   & 13.8  && 19.5& $15.8^{+7.6}_{-5.0}$   & 12.7    \\
10        & 50    & 3096  & $0.72^{+0.33}_{-0.22}$ & 0.69  && 310 & $1.08^{+0.52}_{-0.34}$ & 0.86    \\
10        & 100   & 742  & $1.03^{+0.48}_{-0.32}$ & 0.84  && 197 & $1.25^{+0.61}_{-0.39}$ & 0.98    \\[\cmsTabSkip]
50        & 10    & 2.10  & $125^{+61}_{-39}$      & 102   && 2.38& $72^{+36}_{-23}$       & 58      \\
50        & 50    & 2.57  & $104^{+50}_{-33}$      & 84    && 2.95& $62^{+30}_{-19}$       & 49      \\
50        & 95    & 7.24  & $52^{+25}_{-16}$       & 43    && 10.8& $20.3^{+10.0}_{-6.4}$  & 16.2    \\
50        & 200   & 100   & $2.32^{+1.14}_{-0.73}$ & 1.86  && 84.8& $2.05^{+1.02}_{-0.64}$ & 1.64    \\
50        & 300   & 30.5  & $4.7^{+2.3}_{-1.5}$    & 3.8   && 38.5& $3.7^{+1.9}_{-1.2}$    & 3.0     \\
\end{scotch}
\end{table*}

\begin{table*}[htbp]
  \centering
  \topcaption{Expected and observed event yields, summed over all lepton flavors, in the aggregate signal regions defined by the selection requirements in the table.}
  \label{tab:yields_agg}
\setlength{\tabcolsep}{2pt}
\newcolumntype{y}{D{,}{\,\pm\,}{4.8}}
\begin{scotch}{crcyr}
  Signal region &  $\ptmiss$  (\GeVns{}) &$\mtll$ (\GeVns{})  & \multicolumn{1}{c}{Expected} & Observed \\
  \hline
  A0   & $>$200 & 100--140  & 20.8, 4.4 & 22 \\
  A1   & $>$200 & 140--240  & 6.2,1.0 & 6 \\
  A2   & $>$80  & $>$240      & 1.1,0.4 & 1 \\
\end{scotch}
\end{table*}

\begin{table}[htbp]
  \centering
  \topcaption{Covariance (\cmsLeft) and correlation matrix (\cmsRight) for the background prediction in the aggregate signal regions described in Table~\ref{tab:yields_agg}.}
  \label{tab:cov_agg}
\begin{scotch}{cccc}
\multicolumn{4}{c}{Covariance}\\
& A0 & A1 & A2\\
\cline{2-4}
\multicolumn{1}{r|}{A0}& 13.3 & 1.8 & 0.5\\
\multicolumn{1}{r|}{A1}&      & 0.9 & 0.2\\
\multicolumn{1}{r|}{A2}&      &     & 0.1\\
 \end{scotch}
\ifthenelse{\boolean{cms@external}}{\vspace{2ex}}{\quad}
\begin{scotch}{cccc}
\multicolumn{4}{c}{Correlation}\\
& A0 & A1 & A2\\
 \cline{2-4}
\multicolumn{1}{r|}{A0}& 1    & 0.51 & 0.38\\
\multicolumn{1}{r|}{A1}&      & 1    & 0.49\\
\multicolumn{1}{r|}{A2}&      &      & 1 \\
 \end{scotch}
\end{table}

In order to facilitate the reinterpretation of these results, we construct three aggregate signal regions.
The preselection in Table~\ref{Tab:baselineSel} is applied, but in contrast to the main analysis, there is no separation of events according to lepton flavor.
Regions A0 and A1 are defined as $100\leq\mtll<140\GeV$ and $140\leq\mtll<240\GeV$, with an additional requirement of $\ptmiss>200\GeV$ for both.
Region A2 is defined by $\mtll>240\GeV$ and $\ptmiss>80\GeV$.
Expected and observed yields in the aggregate regions are shown in Table~\ref{tab:yields_agg}.
The covariance and correlation matrices~\cite{Collaboration:2242860} for the background predictions in the aggregate regions are given in Table~\ref{tab:cov_agg}.

\section {Summary}
\label{sec:conclusions}

A search was presented for top squark pair production and dark matter in final states with two leptons, \PQb jets, and large missing transverse momentum
in data corresponding to an integrated luminosity of \lumiGolden in $\Pp\Pp$ collisions collected at a center-of-mass energy of 13\TeV in the CMS detector at the LHC.
An efficient background reduction using dedicated kinematic variables was achieved, suppressing by several orders of magnitude
the large background from standard model dilepton $\ttbar$ events. With no evidence observed for a deviation from the expected background from the standard model,
results were interpreted in several simplified models for supersymmetric top squark pair production as well as through the production
of a spin-0 dark matter mediator in association with a $\ttbar$ pair.

In the \Ttt model with $\PSQtDo \to \PQt\PSGczDo$ decays, $\PSQtDo$ masses $<$800\GeV and $\PSGczDo$ masses $<$360\GeV are excluded.
In the \TbW model with $\PSQtDo \to \PQb\PSGcpDo \to \PQb\PWp\PSGczDo$ decays, $\PSQtDo$ masses $<$750\GeV and $\PSGczDo$ masses $<$320\GeV are excluded, assuming the chargino mass to be the mean of the $\PSQtDo$ and the $\PSGczDo$ masses.
In the newly considered \Tbbllnunu model with decays $\PSQtDo \to \PQb\PSGcpDo \to \PQb\PGn\tilde{\ell} \to \PQb\PGn\ell\PSGczDo$, and therefore 100\% branching to dilepton final states, the sensitivity depends on the intermediate particle masses. With the chargino mass again taken as the mean of the $\PSQtDo$ and the $\PSGczDo$ masses, the strongest exclusion is obtained if the slepton mass is close to the chargino mass.
In this case, excluded masses reach up to 1.3\TeV for $\PSQtDo$ and 800\GeV for $\PSGczDo$.

The \Ttt and \TbW results were combined with complementary searches in the all-jet and single-lepton channels, providing exclusions in
the \Ttt model of $\PSQtDo$ mass $<$1050\GeV for a massless $\PSGczDo$, and a $\PSGczDo$ mass of $<$500\GeV for a $\PSQtDo$ mass of 900\GeV.
In the same way, the \TbW model is excluded for $\PSQtDo$ mass $<$1000\GeV for a massless $\PSGczDo$, and a $\PSGczDo$ mass of $<$450\GeV for a $\PSQtDo$ mass of 900\GeV

The combination extends the sensitivity by $\approx$50\GeV in the masses of both $\PSQtDo$ and $\PSGczDo$ in the \TbW model, and by similar values in the \Ttt model, when the difference between these masses is $\approx$200\GeV.
Aggregate search regions were presented that can be used to reinterpret the results in a wider range of theoretical models of new physics that give rise to the chosen final state.

In addition, the results were interpreted in a simplified model with a dark matter candidate particle coupled to the top quark via a scalar or a pseudoscalar mediator.
Within the assumptions of the model, a scalar mediator with a mass up to 100\GeV and a pseudoscalar mediator with a mass up to 50\GeV are excluded for a dark matter candidate mass of 1\GeV.
The result for the scalar mediator achieves some of the most stringent limits to date in this model.

\begin{acknowledgments}
We congratulate our colleagues in the CERN accelerator departments for the excellent performance of the LHC and thank the technical and administrative staffs at CERN and at other CMS institutes for their contributions to the success of the CMS effort. In addition, we gratefully acknowledge the computing centers and personnel of the Worldwide LHC Computing Grid for delivering so effectively the computing infrastructure essential to our analyses. Finally, we acknowledge the enduring support for the construction and operation of the LHC and the CMS detector provided by the following funding agencies: BMWFW and FWF (Austria); FNRS and FWO (Belgium); CNPq, CAPES, FAPERJ, and FAPESP (Brazil); MES (Bulgaria); CERN; CAS, MoST, and NSFC (China); COLCIENCIAS (Colombia); MSES and CSF (Croatia); RPF (Cyprus); SENESCYT (Ecuador); MoER, ERC IUT, and ERDF (Estonia); Academy of Finland, MEC, and HIP (Finland); CEA and CNRS/IN2P3 (France); BMBF, DFG, and HGF (Germany); GSRT (Greece); OTKA and NIH (Hungary); DAE and DST (India); IPM (Iran); SFI (Ireland); INFN (Italy); MSIP and NRF (Republic of Korea); LAS (Lithuania); MOE and UM (Malaysia); BUAP, CINVESTAV, CONACYT, LNS, SEP, and UASLP-FAI (Mexico); MBIE (New Zealand); PAEC (Pakistan); MSHE and NSC (Poland); FCT (Portugal); JINR (Dubna); MON, RosAtom, RAS, RFBR and RAEP (Russia); MESTD (Serbia); SEIDI, CPAN, PCTI and FEDER (Spain); Swiss Funding Agencies (Switzerland); MST (Taipei); ThEPCenter, IPST, STAR, and NSTDA (Thailand); TUBITAK and TAEK (Turkey); NASU and SFFR (Ukraine); STFC (United Kingdom); DOE and NSF (USA).

\hyphenation{Rachada-pisek} Individuals have received support from the Marie-Curie program and the European Research Council and Horizon 2020 Grant, contract No. 675440 (European Union); the Leventis Foundation; the A. P. Sloan Foundation; the Alexander von Humboldt Foundation; the Belgian Federal Science Policy Office; the Fonds pour la Formation \`a la Recherche dans l'Industrie et dans l'Agriculture (FRIA-Belgium); the Agentschap voor Innovatie door Wetenschap en Technologie (IWT-Belgium); the Ministry of Education, Youth and Sports (MEYS) of the Czech Republic; the Council of Science and Industrial Research, India; the HOMING PLUS program of the Foundation for Polish Science, cofinanced from European Union, Regional Development Fund, the Mobility Plus program of the Ministry of Science and Higher Education, the National Science Center (Poland), contracts Harmonia 2014/14/M/ST2/00428, Opus 2014/13/B/ST2/02543, 2014/15/B/ST2/03998, and 2015/19/B/ST2/02861, Sonata-bis 2012/07/E/ST2/01406; the National Priorities Research Program by Qatar National Research Fund; the Programa Severo Ochoa del Principado de Asturias; the Thalis and Aristeia programs cofinanced by EU-ESF and the Greek NSRF; the Rachadapisek Sompot Fund for Postdoctoral Fellowship, Chulalongkorn University and the Chulalongkorn Academic into Its 2nd Century Project Advancement Project (Thailand); the Welch Foundation, contract C-1845; and the Weston Havens Foundation (USA).
\end{acknowledgments}

\bibliography{auto_generated}

\cleardoublepage \appendix\section{The CMS Collaboration \label{app:collab}}\begin{sloppypar}\hyphenpenalty=5000\widowpenalty=500\clubpenalty=5000\textbf{Yerevan Physics Institute,  Yerevan,  Armenia}\\*[0pt]
A.M.~Sirunyan, A.~Tumasyan
\vskip\cmsinstskip
\textbf{Institut f\"{u}r Hochenergiephysik,  Wien,  Austria}\\*[0pt]
W.~Adam, F.~Ambrogi, E.~Asilar, T.~Bergauer, J.~Brandstetter, E.~Brondolin, M.~Dragicevic, J.~Er\"{o}, M.~Flechl, M.~Friedl, R.~Fr\"{u}hwirth\cmsAuthorMark{1}, V.M.~Ghete, J.~Grossmann, J.~Hrubec, M.~Jeitler\cmsAuthorMark{1}, A.~K\"{o}nig, N.~Krammer, I.~Kr\"{a}tschmer, D.~Liko, T.~Madlener, I.~Mikulec, E.~Pree, N.~Rad, H.~Rohringer, J.~Schieck\cmsAuthorMark{1}, R.~Sch\"{o}fbeck, M.~Spanring, D.~Spitzbart, W.~Waltenberger, J.~Wittmann, C.-E.~Wulz\cmsAuthorMark{1}, M.~Zarucki
% ...
\vskip\cmsinstskip
% ...
1:~~Also at Vienna University of Technology, Vienna, Austria\\
% ...

\end{sloppypar}
\end{document}